%% file: DPF-SiPM.tex
\documentclass[12pt]{article}
\usepackage{graphicx}

\def\gevc{$\rm GeV/c$}
\def\gevc2{$\rm GeV/c^2$}

\def\mevc{$\rm MeV/c$}
\def\mevc2{$\rm MeV/c^2$}

\def\v{$V_{\rm b}$}

\def\t{$T$}
\def\g{$G$}
\def\dgdt{$dG/dT$}
\def\dgdv{$dG/dV$}
\def\dvdt{$dV_{\rm b}/dT$}

\input econfmacros.tex
\def\Title#1{\begin{center} {\Large {\bf #1} } \end{center}}

\begin{document}

\Title{Gain Stablization of SiPMs}

\bigskip\bigskip


\begin{raggedright}  

{\it Gerald Eigen\index{Eigen, G.}\footnote{on behalf of G. Eigen, A. Tr¾et, J. Zalieckas (Bergen U.), J. Cvach, .i Kvasnicka, I. Polak (Prague, Inst. Phys., CAS)}\\
Department of Physics and Technology\\
University of Bergen\\
N-5007 Bergen,  NORWAY
}
\bigskip\bigskip

Talk presented at the APS Division of Particles and Fields Meeting (DPF 2017), July 31-August 4, 2017, Fermilab. C170731.

\abstract{The gain of silicon photomultipliers (SiPMs) increases with bias voltage and decreases with temperature. To operate SiPMs at stable gain, the bias voltage can be adjusted to compensate temperature changes. We have tested this concept with 30 SiPMs from three manufacturers (Hamamatsu, KETEK, CPTA) in a climate chamber at CERN varying the temperature from $1^\circ \rm C$ to $50^\circ \rm C$. We built an adaptive power supply that used a linear temperature dependence of the bias voltage readjustment. With one selected bias voltage readjustment, we stabilized four SiPMs simultaneously. We fulfilled our goal of limiting the deviation from gain stability in the $20^\circ \rm C-30^\circ C$ temperature range to less than $\pm 0.5\%$ for most of the tested SiPMs. We have studied afterpulsing of SiPMs for different temperatures and bias voltages.}

\end{raggedright}

\section{Introduction}

\begin{figure}[htbp!]
\centering
\includegraphics[width=75mm]{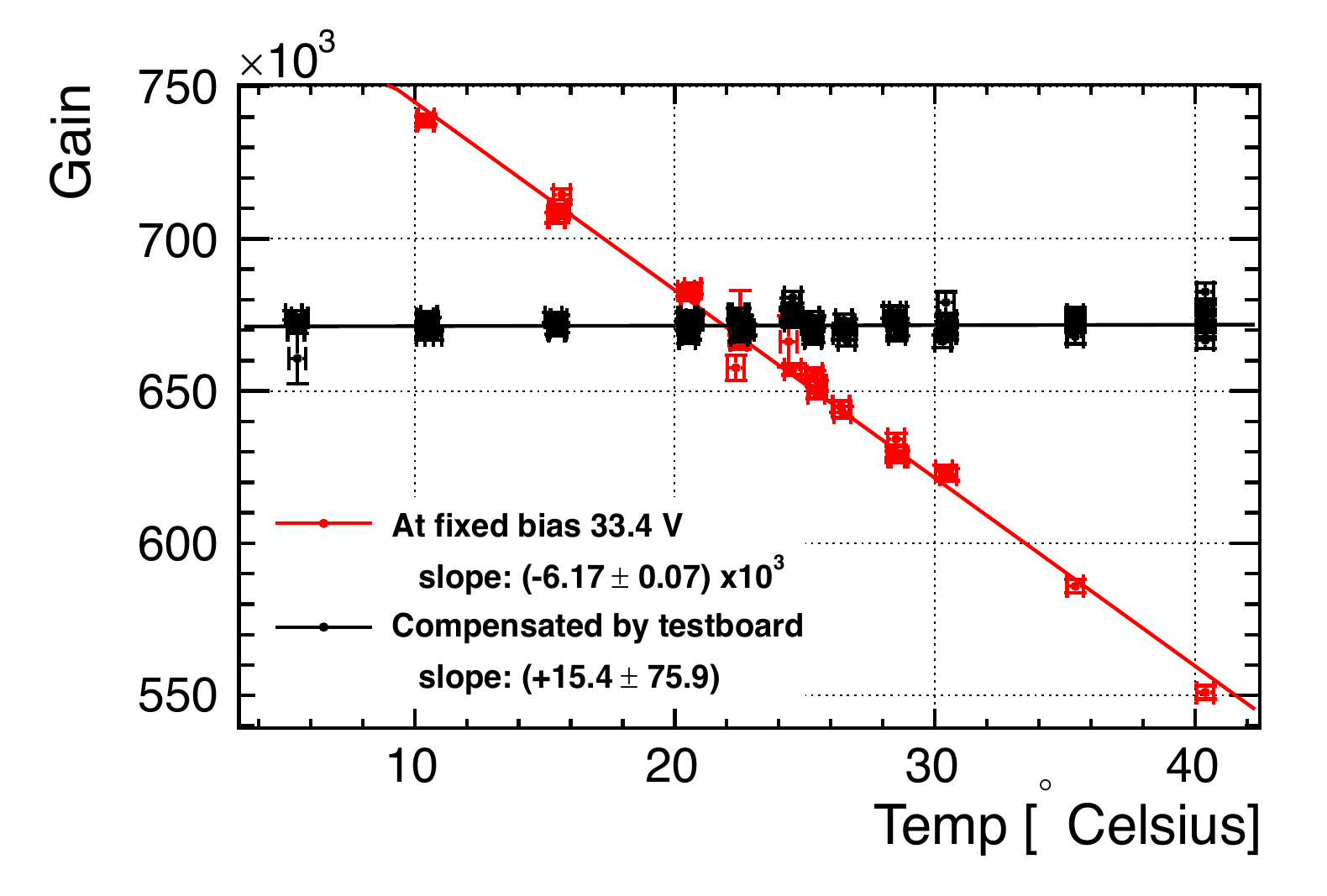}
\caption{The temperature dependence of the gain of SiPMs without \v\ adjustments (red curve) and with  \v\ adjustments (black curve).}
\label{fig:uncorrected-gain} 
\end{figure}

The gain \g\ of silicon photomultipliers (SiPMs)~\cite{Bondarenko, Buzhan2002,Buzhan2003} increases with bias voltage \v\ and decreases with temperature \t. To operate SiPMs at stable gain, the bias voltage can be adjusted to compensate for temperature changes. 
This is particularly important for the operation of large detector system like analog hadron calorimeter~\cite{AHCAL}. This task requires the knowledge of \dvdt\, which is obtained from measurements of \dgdv\ and \dgdt. Figure~\ref{fig:uncorrected-gain} shows the dependence of the 
gain on temperature without and with \v\ adjustments. We define stable gain if  the gain change $\Delta G$ satisfies the condition $\Delta G/G <\pm 0.5\%$ in the $20-30^\circ$C. We use an adaptive power supply that accomplishes automatic \dvdt\ adjustments when the temperature changes. The \dvdt\ correction uses a linear approximation. 

\section{Experimental Setup}

\begin{figure}[htbp!]
\centering 
\includegraphics[width=90 mm]{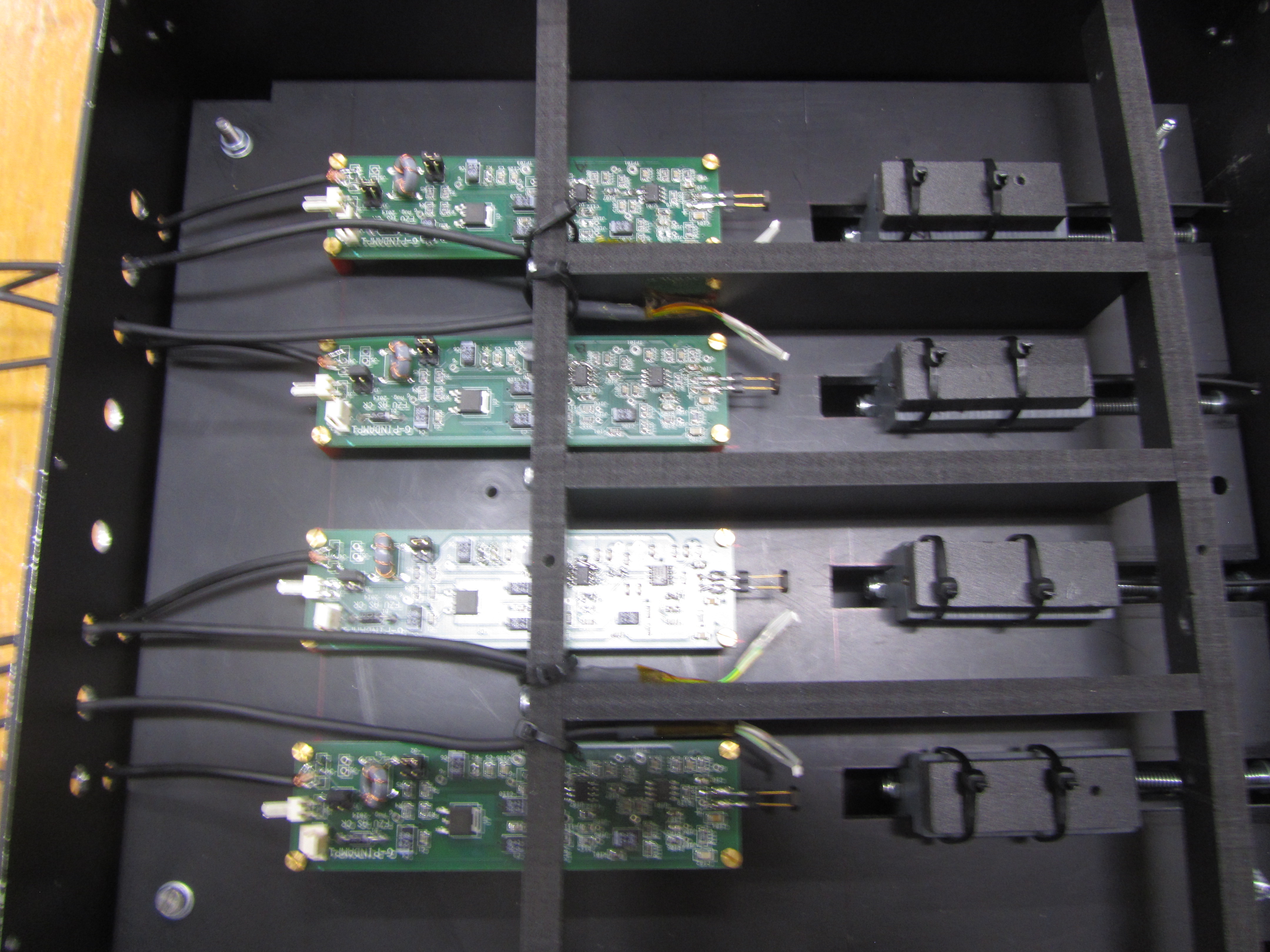}
\caption{ Setup of the gain stabilization measurements inside the black box. The green circuit boards host the preamplifiers and signal readout. The black cable attached is the signal cable. Each SiPM is inserted into a connector on the left-hand side. The Pt1000 sensors positioned near the SiPM are clearly visible. The clear fibers transporting the blue LED light run inside the black distance-adjustable foam boxes on the right, which were  mounted precisely to illuminate the SiPM uniformly.}
\label{fig:setup}
\end{figure}

We performed gain stabilization studies  in a climate chamber at CERN. Figure~\ref{fig:setup} shows the experimental setup, which an improvement of a previous study~\cite{eigen}. Four SiPMs are tested simultaneously. They are housed in separate compartments inside

\begin{figure}[htbp!]
\centering 
\includegraphics[width=70 mm]{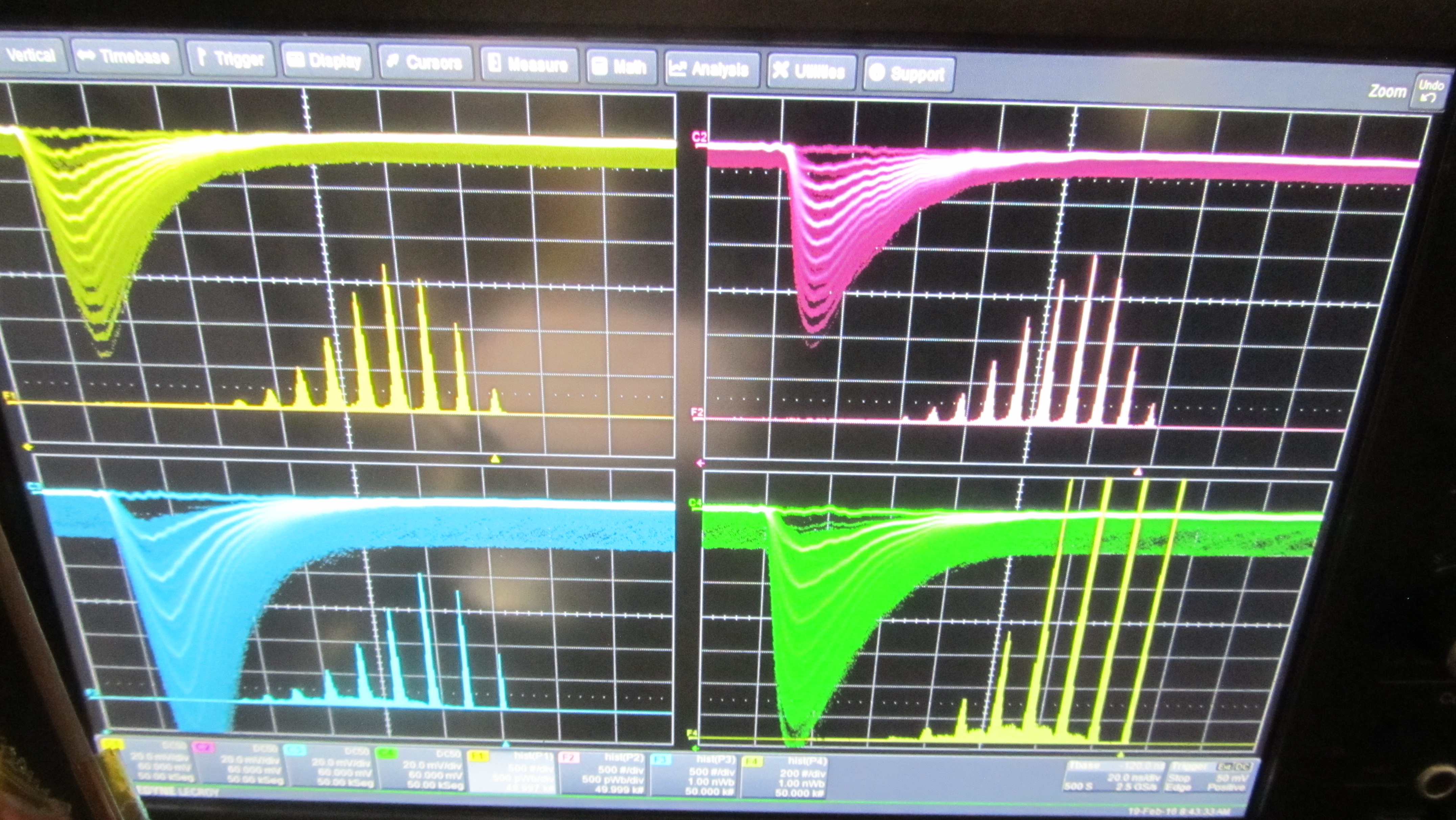}
\includegraphics[width=77 mm]{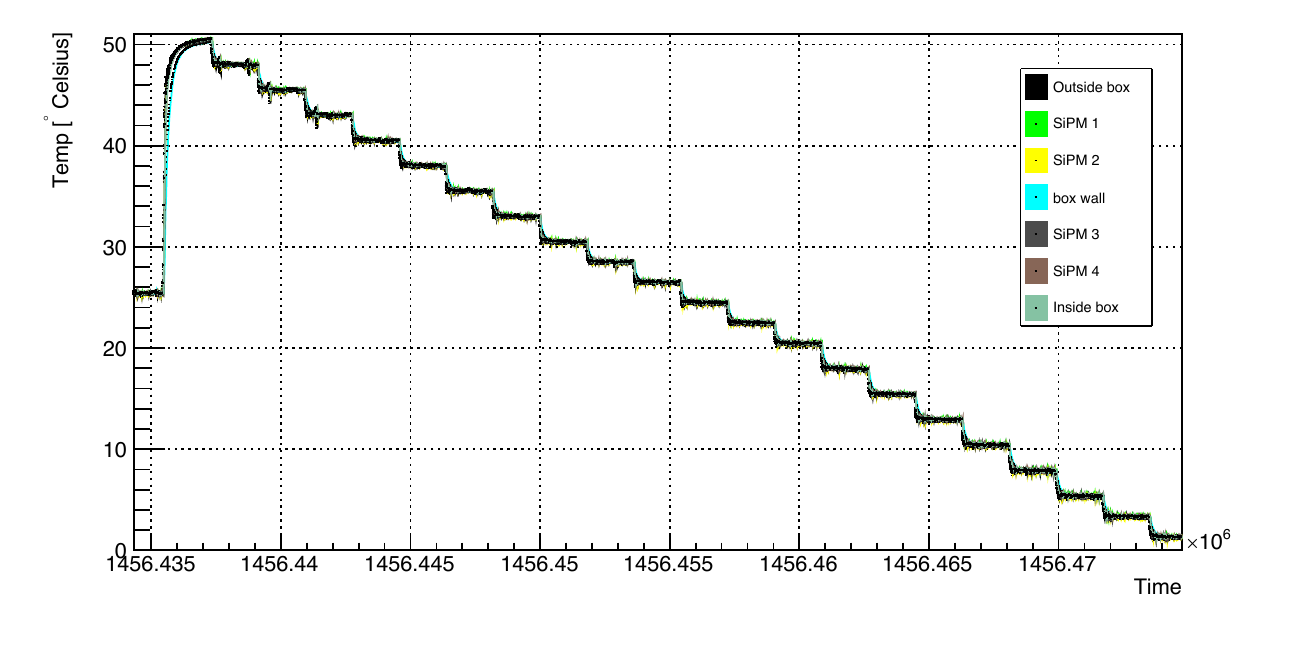}
\caption{Waveform and photoelectron spectra of four Hamamatsu S13360  MPPCs (left). Typical temperature profile for an overnight stabilization run
(right).  }
\label{fig:waveform}
\end{figure}

\noindent
a black box to prevent optical cross talk. Each SiPM is read out with a two-stage preamplifier.  The amplified signals are recorded by the four channels of a digital oscilloscope that uses 12-bit ADCs (LeCroy HDO6104). We illuminate each SiPM with blue LED light. To minimize noise pickup, the LED is placed outside the climate chamber and the light is transported via clear fibers, which are positioned by the distance-adjustable black foam boxes. The LED is trigged by a  3.4 ns wide light pulser signal. Both repetition rate and intensity are adjustable. We operate the light pulser with a rate of 10 kHz. The intensity is adjusted such  that  several photoelectron peaks are produced in addition to a clearly visible pedestal. We record the waveforms of the four channels from the digital oscilloscope with a 5 Mb/s sampling rate. For example, Fig.~\ref{fig:waveform} (left) shows  50000 recorded waveforms and the resulting photoelectron spectra for Hamamatsu MPPCs with trenches (S13360). Individual photoelectron peaks are clearly separated.  
 We use seven temperature sensors to record the temperature accurately inside the climate chamber. As seen in Fig.~\ref{fig:setup}, one sensor is placed near each SiPM, a fifth sensor is placed inside the black box, a sixth sensor is attached to the outside wall of the black box, and the seventh sensor is placed in the climate chamber outside the black box.  
 
We have tested gain stabilization for 30 SiPMs, 18 from Hamamatsu, eight from KETEK and four from  CPTA. Table~\ref{tab:sipm} in Appendix A summarizes properties of these photodetectors.  For Hamamatsu and KETEK SiPMs, the LED light impinges directly on the surface of the photodetector. The CPTA sensors were glued to a wavelength-shifting fiber inserted into a groove in a scintillator tile. Thus, the light passed through the scintillator and typically was absorbed by the wavelength-shifting fiber that reemitted it at higher wavelength and transported it to the SiPM (see section~\ref{CPTA-study} and Fig.~\ref{fig:tile-RO}). 
The gain stabilization study consists of two steps. In the first step we determine $dV_{\rm b}/dT$ from $dG/dV_{\rm b}$ and $dG/dT$ measurements for each SiPM. In the  second step we select a common value of $dV_{\rm b}/dT$ to test gain stabilization of four SiPMs simultaneously. 
Figure~\ref{fig:waveform} (right) shows a typical temperature profile used for the gain stabilization measurements.

\subsection{Extraction of Photoelectron Spectra}

\begin{figure}[htbp!]
\centering 
\includegraphics[width=140 mm]{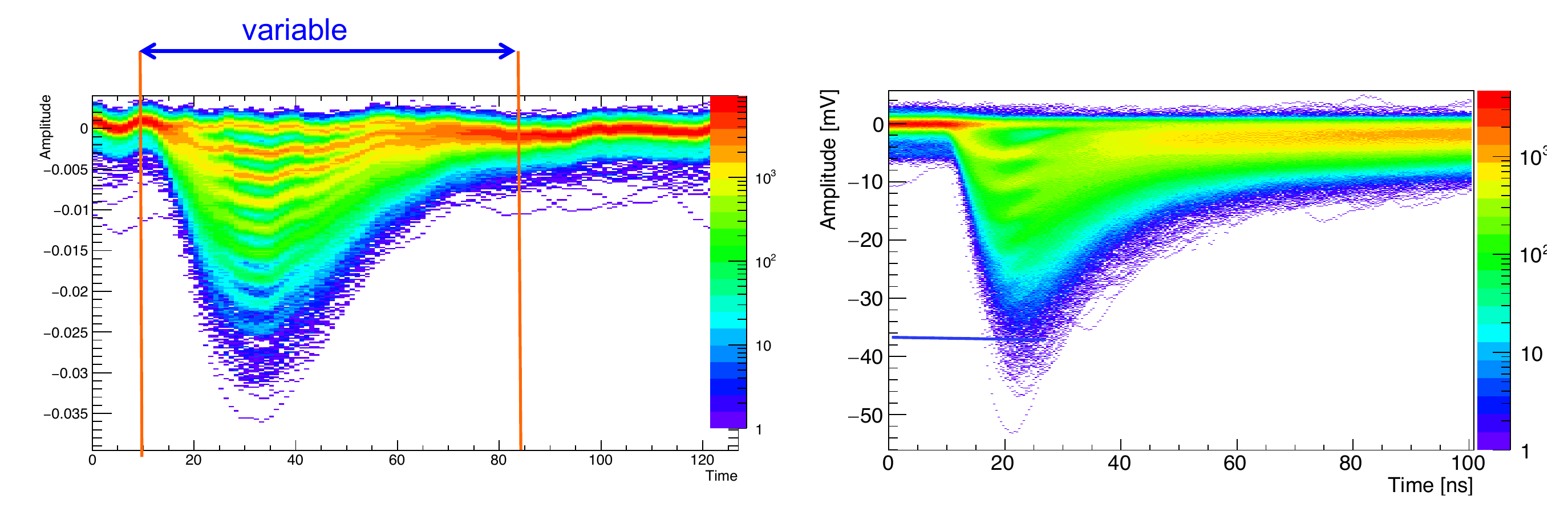}
\includegraphics[width=140 mm]{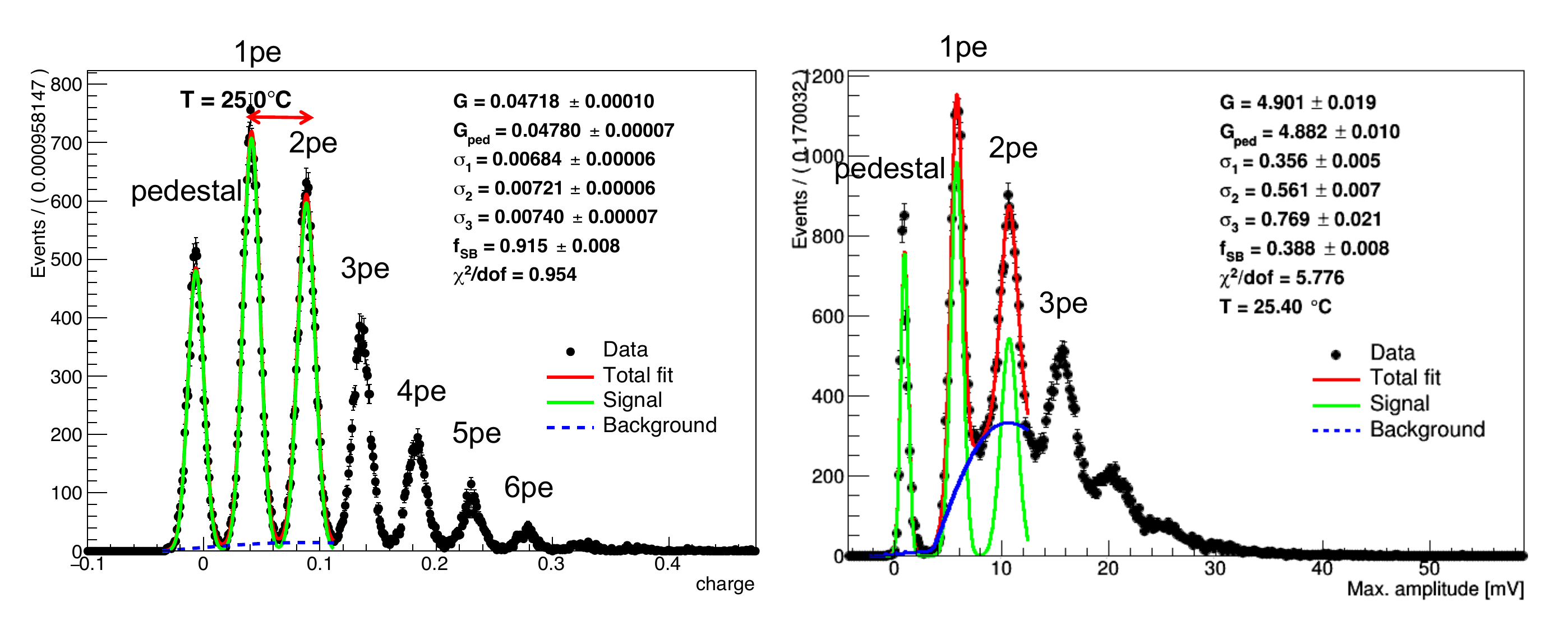}
\caption{Waveform of a Hamamatsu MPPC without trenches (top left) and waveform of a CPTA SiPM (top right). Photoelectron spectra  resulting from integrating the waveform above over a variable time window (bottom left) and from the minimum value of the waveform above (bottom right). The red arrow shows the gain.}
\label{fig:pe}
\end{figure}

First, we need to convert the waveforms into photoelectron spectra. For all Hamamatsu MPPCs, we integrate  each waveform over a variable time window. The starting point is given by the starting point of the waveform while the length is determined by the time the waveform reaches the baseline again. 
From studies with fixed and variable integration windows we concluded that the variable integration window gave the best performance.
Figures~\ref{fig:pe} (top left) and ~\ref{fig:pe} (bottom left) show the waveform and photoelectron spectrum that is obtained with this procedure. 
For KETEK and
CPTA SiPMs, we extract the photoelectron spectrum from the waveform minimum since for certain \v\ and \t\ the integration over the waveform does not produce well separated photoelectron peaks. Figures~\ref{fig:pe}  (top right) and ~\ref{fig:pe} (bottom right) show the waveform and photoelectron spectrum of a CPTA sensor, which shows clearly separated photoelectron peaks. 
The gain is defined as the distance between first and second photoelectron peaks. This choice yields more robust results than choosing the distance between pedestal and first photoelectron peak.

\subsection{Analysis Methodology}

\begin{figure}[htbp!]
\centering 
\includegraphics[width=140 mm]{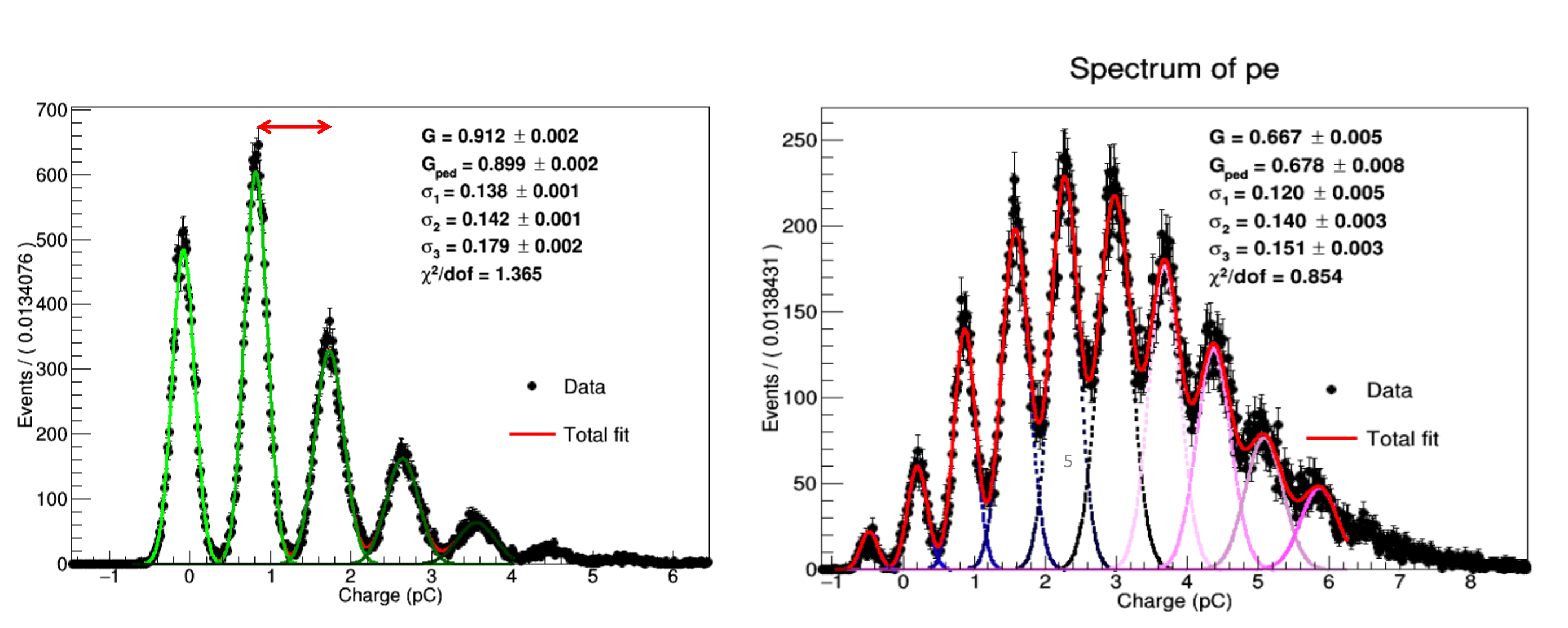}
\caption{Photoelectron spectrum with the fit of the original fit model  overlaid  (left) and photoelectron spectrum with the fit of the new signal model overlaid (right). The red arrow show the gain. }
\label{fig:model}
\end{figure}

To determine the gain of SiPMs,  we fit the photoelectron spectra with the likelihood function
\begin{equation}
{\cal L} = \prod_{i=1}^{50000} [ f_{s} F_{sig}(w_i) + (1-f_{s} )F_{bkg}(w_i) ]
\end{equation}
where $f_{s} $ is the signal fraction and $F_{sig}(w^\prime)$ and $F_{bkg}(w^\prime)$ are the signal and background probability density functions (pdf), respectively.
In our original fit methodology, the signal pdf  consists of a Gaussian function ($G_{ped}(Q)$) for the pedestal plus two Gaussian functions ($G_1(Q), ~G_2(Q)$) for the first  and second photoelectron peaks
\begin{equation}
F_{sig}(w_i)=f_{ped} G_{ped}(w_i) + f_1 G_1(w_i) + (1-f_{ped} -f_1)G_2(w_i)
\end{equation}
where $f_{ped}$ and $f_1$ represent the fractions of the corresponding Gaussian functions. The background pdf ($F_{bkg}(w_i)$ is determined by a sensitive nonlinear iterative peak-clipping algorithm (SNIP) available in ROOT~\cite{Root}. Positions, widths  and fractions of the Gaussian functions are free parameters in the fit.
In the new fit methodology, we fit the pedestal and all visible photoelectron peaks with Gaussians ($G_{ped}(Q)$ and $G_i (Q)$), where all widths and fractions are free parameters
\begin{equation}
F_{sig}(w_i)=f_{ped} G_{ped} (w_i)+ \sum_{j=1}^{n-1} f_j G_j (w_i)+ (1-f_{ped} - \sum_{j=1}^{n-1}f_j)G_n(w_i).
\end{equation}
For Hamamatsu MPPCs without trenches no background pdf is needed, Similarly, for  KETEK and CPTA SiPMs so far fitted with the new model, no background pdf is necessary. However, Hamamatsu MPPCs with trenches that show tails on the right-hand side of each photoelectron peak need at least a modified signal pdf.
Figure~\ref{fig:model} (left) shows the photoelectron spectrum with the fit result of our original fit model overlaid while Fig.~\ref{fig:model} (right) shows a photoelectron spectrum with the fit result of the new model overlaid. 
Using binned fits, we have fitted all photoelectron spectra of all SiPMs with the original model. Concerning the new fit model, we have fitted so far only Hamamatsu MPPCs with trenches as well as one KETEK and one CPTA SiPM. 

\section{Determination of \dvdt}

\begin{figure}[htbp!]
\centering 
\includegraphics[width=140 mm]{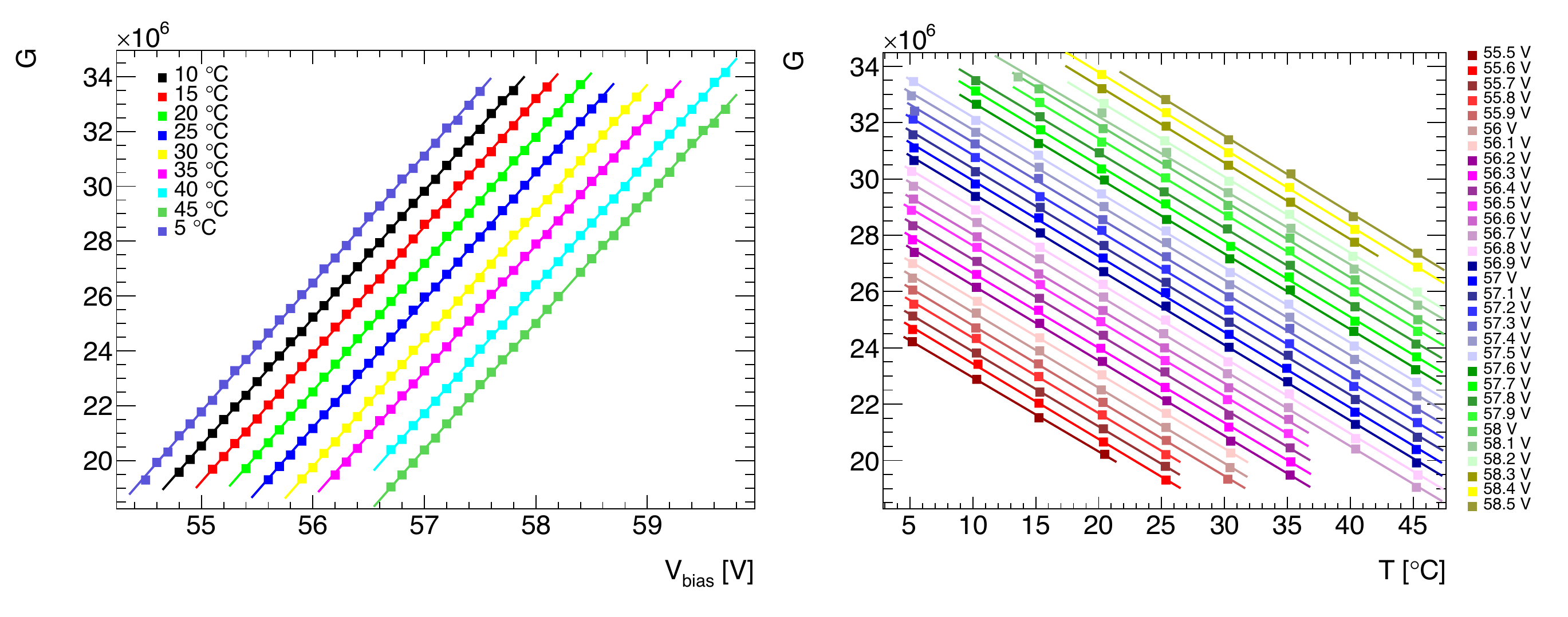}
\caption{Measurements of gain versus bias voltage for different temperatures (left) and gain versus  temperature for different  bias voltages (right) for the Hamamatsu MPPC S13360-1325.}
\label{fig:GvV}
\end{figure}

\begin{figure}
\centering 
\includegraphics[width=150 mm]{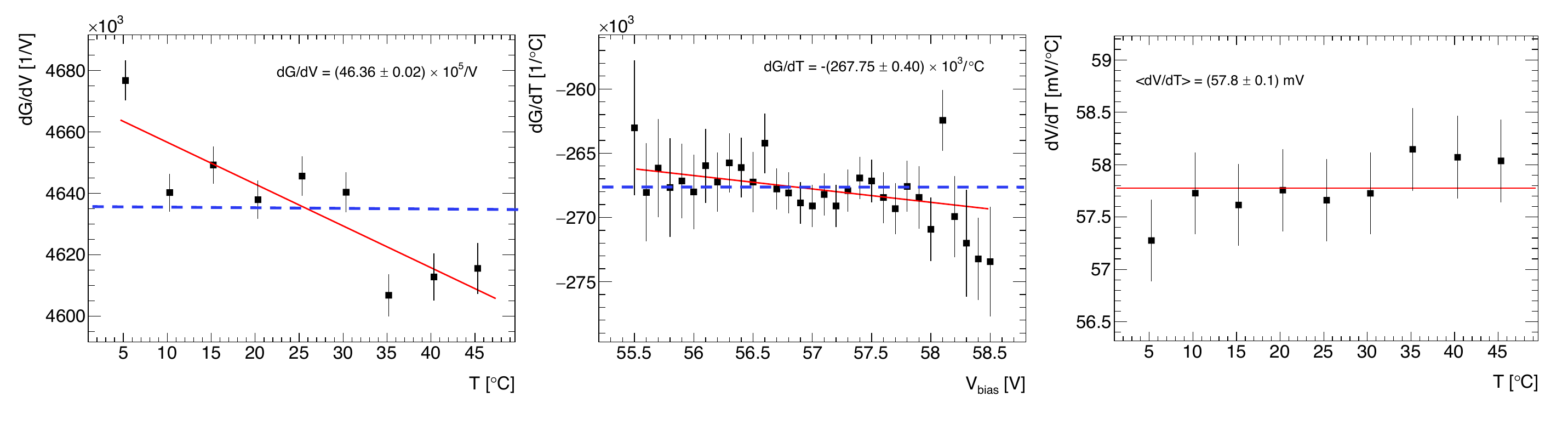}
\caption{Measurements of the \dgdv\ slopes versus temperature (left), \dgdt\ versus bias voltage (middle) and \dvdt\ versus temperature (right) 
for the Hamamatsu MPPC S13360-1325.}
\label{fig:dgdv}
\end{figure}

For each SiPM, we first measure the gain versus bias voltage for different temperatures. Figure~\ref{fig:GvV} (left) shows the results  for the Hamamatsu MPPC S13360-1325.
For each temperature, all gain measurements are fitted with a linear function. We plot the resulting \dgdv\ slopes as a function of temperature in Fig.~\ref{fig:dgdv} (left).  A fit with
a linear function yields $dG/dV_{\rm b} =(4.636\pm 0.002_{\rm stat})\times 10^6/ \rm V$ for the constant at $T=25^\circ$C. The error is statistical. The observed deviation from uniformity is small and amounts to a linear dependence of  $\pm 0.5\%$ between $5^\circ$ and $40^\circ$C. Figure~\ref{fig:GvV} (right) shows the gain  versus  temperature  for different bias voltage. For each bias voltage we fit all gain measurements with a linear function. 
We plot the resulting \dgdt\ slopes 

\noindent
as a function of bias voltage in Fig.~\ref{fig:dgdv} (middle).  A fit with a linear function yields 
$dG/dT = -(2.678 \pm 0.004_{\rm stat}) \times 10^6/ \rm ^\circ C$ for the constant at the nominal bias voltage. The observed deviation from uniformity is small and amounts to a linear dependence of $\pm 0.65\%$ in the $55.5~\rm V$ to $58.5~\rm V$ range. We average all \dgdt\ slopes and divide them by \dgdv. Figure~\ref{fig:dgdv} (right) shows the results as a function of temperature. A fit to a uniform distribution yields $dV_{\rm b}/dT =(57.8\pm 0.1)~\rm mV/^\circ C$. The error results from the rms from averaging individual $dV_{\rm b}/dT$ measurements. 
We have performed these measurements for 30 SiPMs. 
Table~\ref{tab:stabilization} lists the measured $dV_{\rm b}/dT$ values for all 30 SiPMs. 

\begin{figure}
\centering 
\includegraphics[width=140 mm]{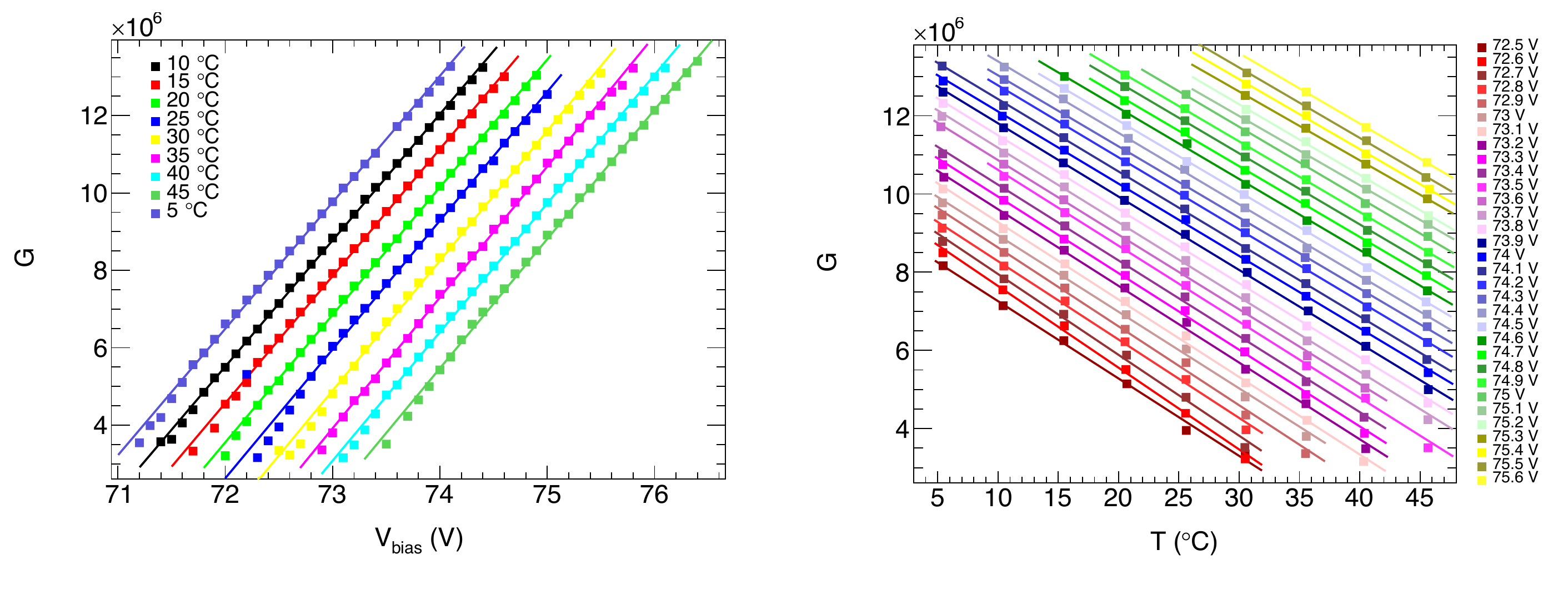}
\caption{Measurements of gain versus bias voltage for different temperatures (left) and gain versus  temperature for different  bias voltages (right) for the Hamamatsu MPPC S13360-1325.}
\label{fig:GvVa}
\end{figure}

\begin{figure}
\centering 
\includegraphics[width=150 mm]{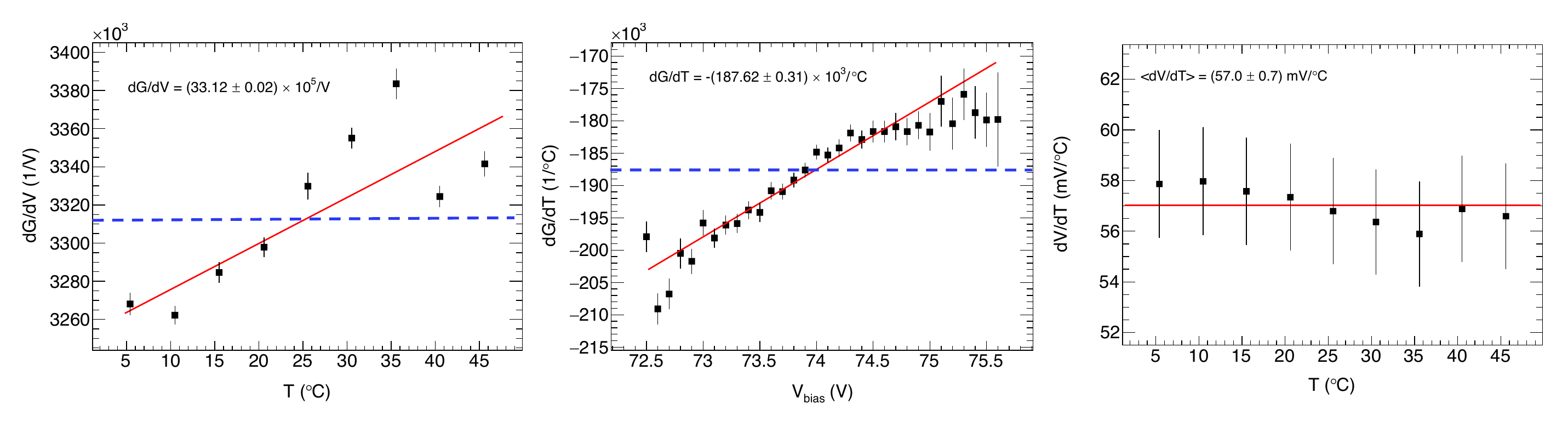}
\caption{Measurements of the \dgdv\ slopes versus temperature (left), \dgdt\ versus bias voltage (middle) and \dvdt\ versus temperature (right) 
for the Hamamatsu MPPC S13360-1325.}
\label{fig:dgdva}
\end{figure}

To ascertain that our results are stable with respect to the fit methodology, we started to determine \dgdv, \dgdt\ and \dvdt\ with the new fit methodology. We have fitted all Hamamatsu MPPC without trenches as well as one KETEK and one CPTA SiPM.  Figures~\ref{fig:GvVa} and~\ref{fig:dgdva} show the corresponding plots for the new fit methodology for the Hamamatsu MPPC B2-20. We determine $dG/dV_{\rm b} = (33.12\pm 0.02_{\rm stat}) \times 10^6/\rm V$, $dG/dT =-(1.876\pm 0.003_{\rm stat}) \times 10^5/^\circ \rm C$, and $dV_{\rm b}/dT =(57.0 \pm 0.1)~\rm  mV/^\circ  C$. Again, the latter error results from the rms from averaging individual $dV_{\rm b}/dT$ measurements.

We show a direct comparison of the two fitting methodologies for the Hamamatsu MPPC B2-20, the CPTA SiPM 1065 and the KETEK SiPM $\rm PM3350_6$. Figure~\ref{fig:comparefits} in appendix A shows the $dV_{\rm b}/dT$ measurements for these three SiPMs and Table~\ref{tab:comparefits} in appendix A shows the measured $dV_{\rm b}/dT$ values.

We obtain consistent $ dV_{\rm b}/dT$ results with the two fit methodologies. For 12 Hamamatsu MPPC, all $dV_{\rm b}/dT$ results are within errors. 
For KETEK and CPTA SIPMs we have tested the new fitting methodology only on one  channel so far. 
For these two SiPMs, $dV_{\rm b}/dT$  values agree within two  standard deviations. 
We will fit the remaining KETEK and CPTA SiPMs. For Hamamatsu S13360 and LCT MPPCS, however, we need an additional  function to represent a small tail at the right-hand side on each   photoelectron peak.

\section{Gain Stabilization Studies}

Before each overnight stabilization run, we determined $dV_{\rm b}/dT$ as well as possible for the four SiPMs  to be tested. We selected a value for the stabilization that was suitable to stabilize these four SiPMs simultaneously. We select the operating bias voltage at a given temperature and the $dV_{\rm b}/dT$ slope on the bias voltage regulator board~\cite{cvach}. We typically tested gain stability within the range $1^\circ \leq T \leq 48^\circ$C taking at least 18 sets of 50000 waveform  samples of at a selected temperature value. We used only measurements for which the newly selected temperature was constant. With this procedure we tested all 30 SiPMs.

\subsection{ Studies of Hamamatsu MPPCs}

We tested 18 MPPCs from Hamamatsu: eight experimental devices and four commercially available devices with no trenches plus two experimental devices and four commercially available devices with trenches. The trenches reduce the pixel-to-pixel cross talk. For the stabilization test we select similar-type sensors. The nominal bias voltage for SIPMs without trenches lies around 65-75~V while that for MPPC with trenches lies around 50-60~V. 
More details on the MPPC are given Tab.~\ref{tab:sipm} in appendix A. 


\begin{figure}[htbp!]
\centering 
\includegraphics[width=140 mm]{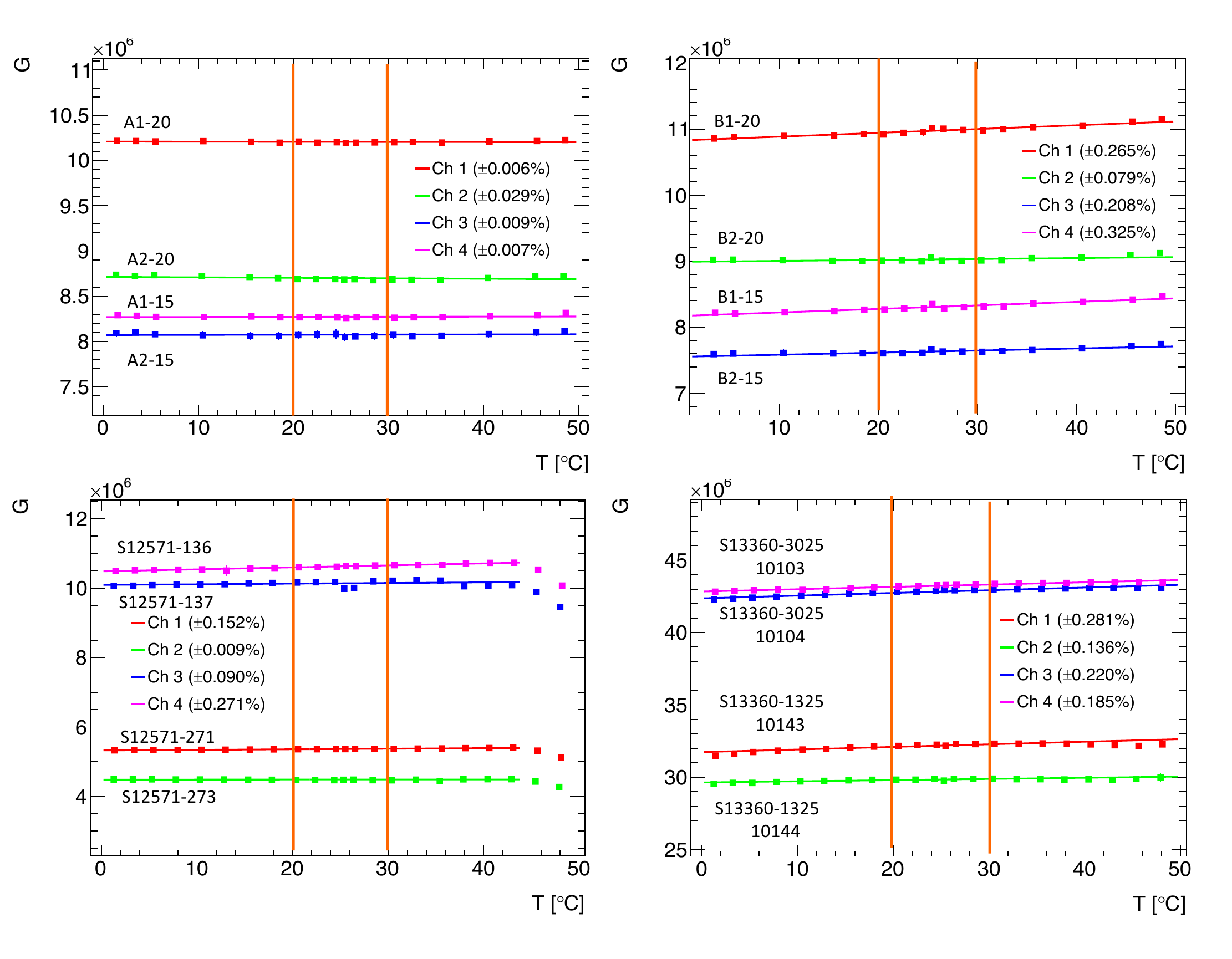}
\caption{Measurements of gain versus temperature after stabilization  for Hamamatsu MPPCs extracting the gain with the original fit methodology:  A-sensors (top left), B-sensors (top right), S12571 sensors (bottom left) and S13360 sensors (bottom right). The vertical bars show the temperature range of interest.}
\label{fig:H-stab}
\end{figure}
\begin{figure}
\centering 
\includegraphics[width=140 mm]{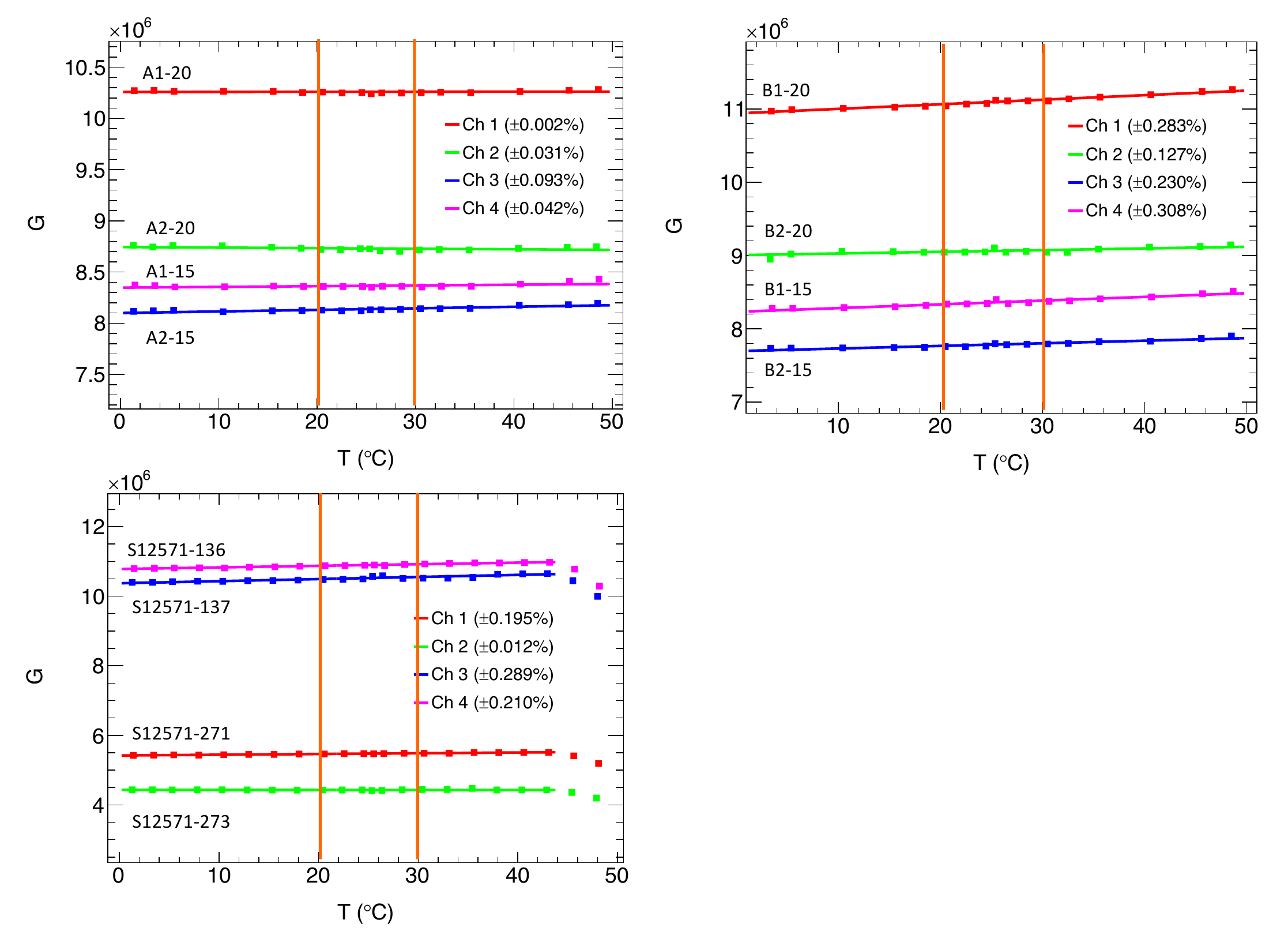}
\caption{Measurements of gain versus temperature after stabilization  for Hamamatsu MPPCs extracting the gain with the new fit methodology:  A-sensors (top left), B-sensors (top right) and S12571 sensors (bottom left). The vertical bars show the temperature range of interest.}
\label{fig:H-stabn}
\end{figure}

\begin{figure}
\centering 
\includegraphics[width=130 mm]{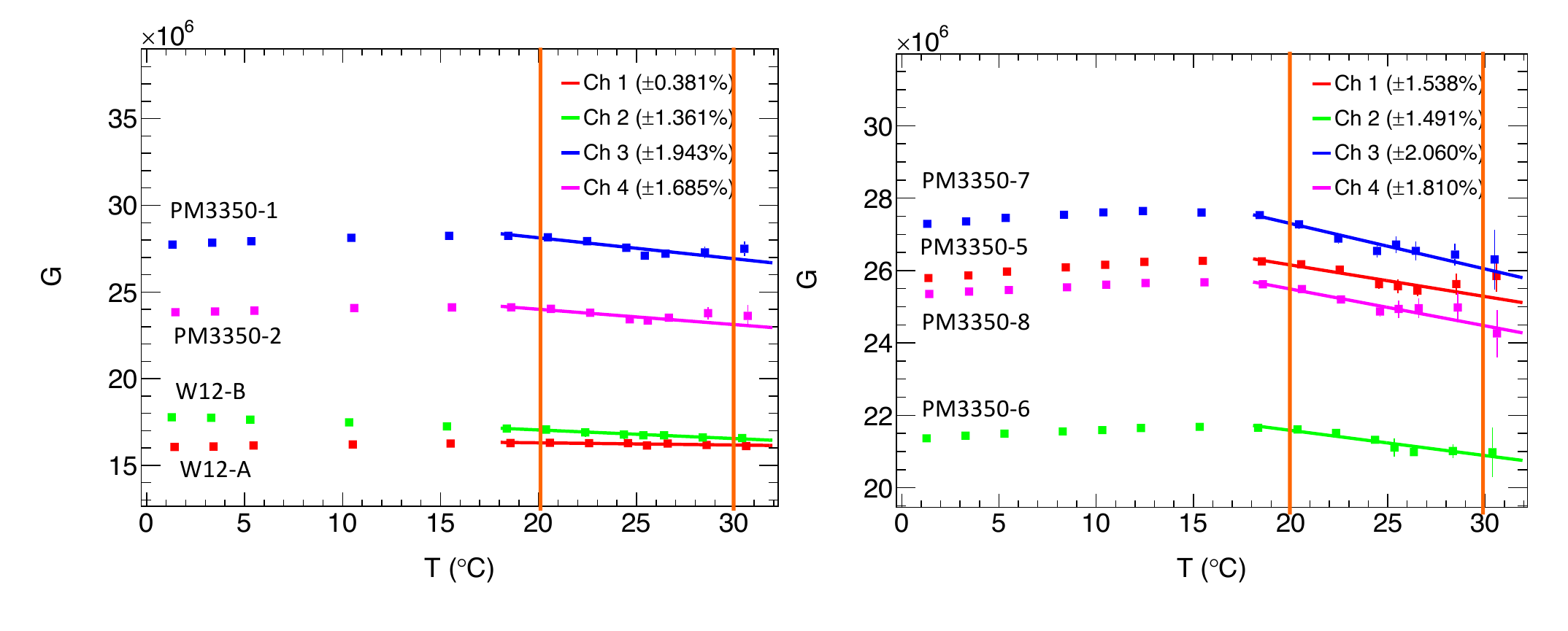}
\caption{Measurements of gain versus temperature after stabilization  for KETEK SiPMs extracting the gain with the original fit methodology: W12 and two PM3350 sensors (left) and four PM3350 sensors  (right). The vertical bars show the temperature range of interest.}
\label{fig:K-stab}
\end{figure}

Figure~\ref{fig:H-stab} shows the results of the gain stabilization for Hamamatsu MPPCs A-sensors, B-sensors, S12571 sensors and S13360 sensors using the original fit methodology.  Table ~\ref{tab:non-uniformity} summarizes our results of linear deviation from uniformity in the  $20^\circ -30^\circ$C. All Hamamatsu MPPCs satisfy our criteria of $\Delta G/G<±0.5\% $ in the $20^\circ -30^\circ$C  temperature range. Most of the MPPCs actually satisfy this criterion in the fully tested   temperature range ($1^\circ -45^\circ$C). Figure~\ref{fig:H-stabn} shows the corresponding results of the gain stabilization for Hamamatsu MPPCs A-sensors, B-sensors and  S12571  sensors for the new fit methodology. The results are also listed in Tab.~\ref{tab:non-uniformity} and are consistent with those obtained with the original fit methodology.

\subsection{ Studies of KETEK SiPMs}
 
We tested eight SiPMs from KETEK, six commercially available photodetectors PM3350 and two experimental devices W12. 
The nominal bias voltage is around 28~V. Further details are given in Table~\ref{tab:sipm} in appendix A. 
The decay time of KETEK SiPM waveform is much longer than that of other SiPMs. Within the 200~ns integration window the waveforms typically do not return to the baseline. We therefore  extract the SiPM photoelectron spectrum from the minimum position of the waveform. The SiPMs do not work properly  at temperatures above $30^\circ$C. 
 
We performed gain stabilization of the KETEK SiPMs in two batches with $dV_{\rm b}/dT =18.2~\rm mV/^\circ C$. Figure~\ref{fig:K-stab} shows the gain versus temperature after stabilization.  The KETEK sensors show a more complicated $V(T)$ dependence. A linear gain compensation is not sufficient. For low temperatures ($\rm 1^\circ - 18^\circ C$) the gain rises slowly remaining constant in the range $\rm 18^\circ - 22^\circ C$ before declining again. Out of the eight SiPMs only W12-A satisfies our criterion of $\Delta G/G < 0.5\%$ in the $\rm 20^\circ - 30^\circ C$ temperature range. Table ~\ref{tab:non-uniformity} summarizes our results of linear deviation from uniformity in the  $20^\circ -30^\circ$C.

\subsection{ Studies of CPTA SiPMs}
\label{CPTA-study}

\begin{figure}
\centering 
\includegraphics[width=60 mm]{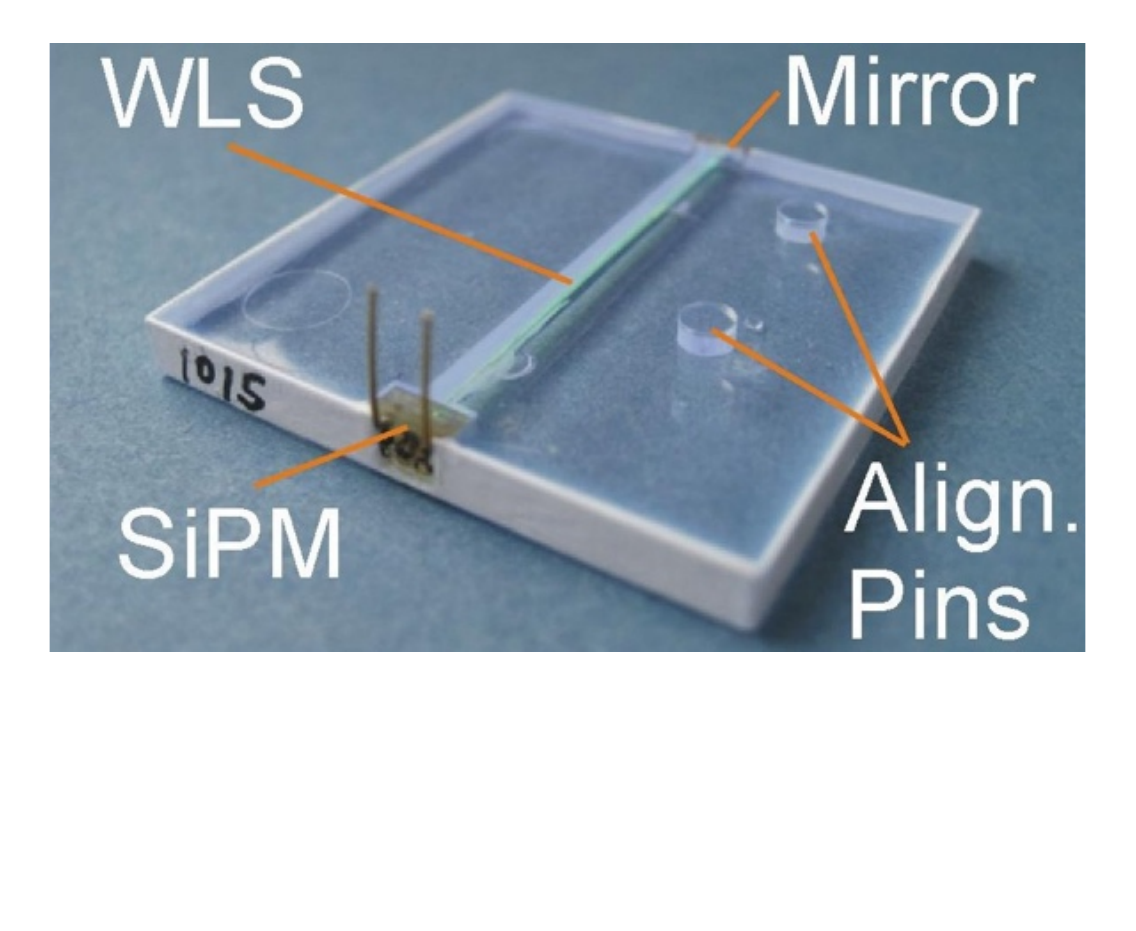}
\includegraphics[width=75 mm]{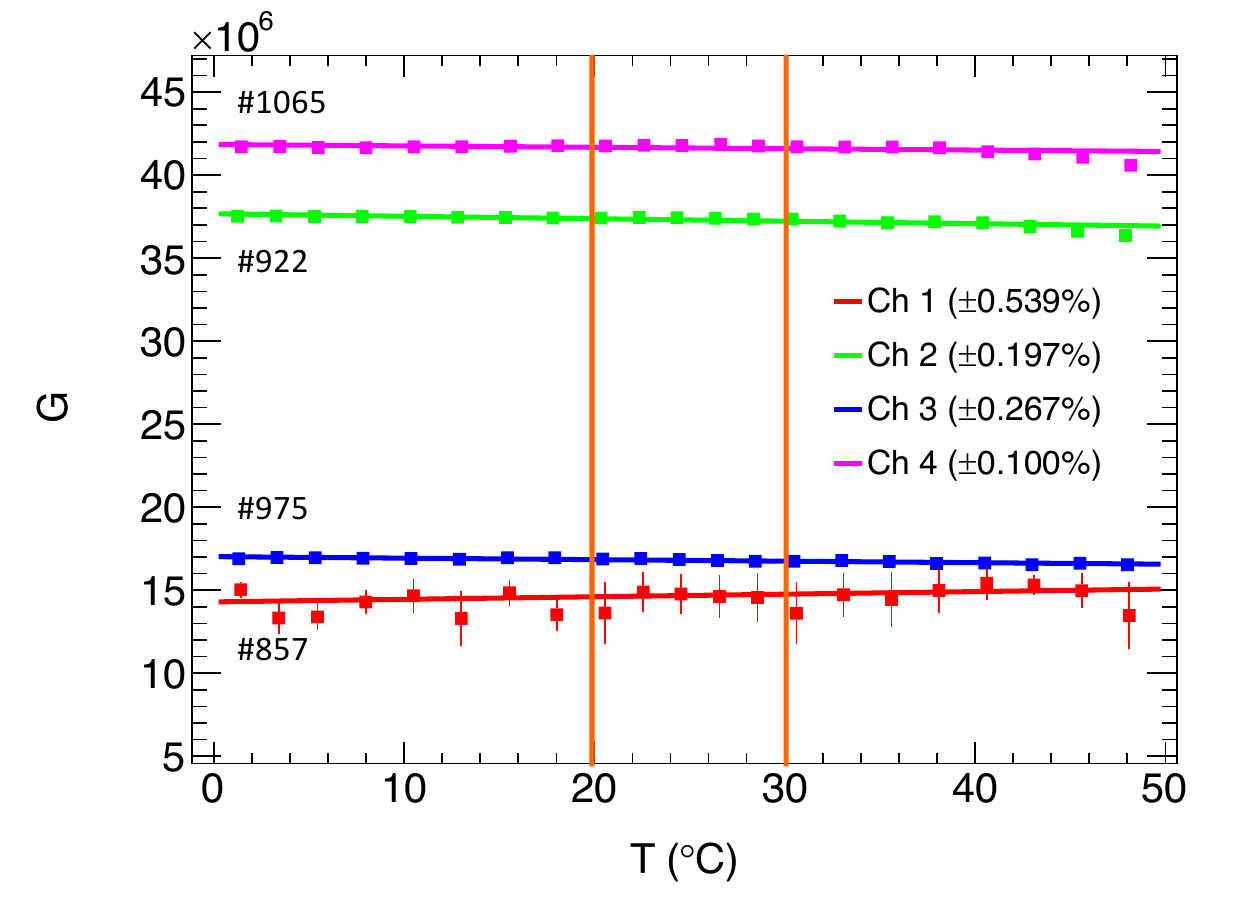}
\caption{Mounting of CPTA SiPM to wavelength-shifting fiber inserted into a groove in a scintillator tile (left). Measurements of gain versus temperature after stabilization  for CPTA SiPMs extracting the gain with the original fit methodology (right). The vertical bars show the temperature range of interest }
\label{fig:tile-RO}
\end{figure}

The CPTA SiPMs have been glued to a wavelength-shifting fiber inserted into a groove in a 3~mm thick $\rm 3 ~cm \times 3~cm$ scintillator tile as Fig.~\ref{fig:tile-RO} (left) shows. So the illumination proceeded via the scintillator tile and not directly onto the SiPM. We tested four such configurations. 
Further details are given in Table~\ref{tab:sipm} in appendix A

We used $dV_{\rm b}/dT=\rm 21.2~ mV/^\circ C$ to stabilize the four SiPMs  simultaneously in the  $\rm 1^\circ - 48^\circ C$ temperature range. Figure~\ref{fig:tile-RO} (right) shows the results of the gain stabilization for the CPTA sensors. The gain is nearly uniform up to $\rm 30^\circ C$.  The operation of SiPMs \#922 and \#1065 look fine; SiPM \#857 was rather noisy and SiPM \#975 changed gain after at $T=\rm 45^\circ C$ but worked fine afterwards. Three CPTA SiPMs satisfy our requirement of $\Delta G/G < 0.5\%$ in the $\rm 20^\circ - 30^\circ C$ temperature range. Table ~\ref{tab:non-uniformity} summarizes our results of linear deviation from uniformity in the  $20^\circ -30^\circ$C.

\begin{figure}
\centering 
\includegraphics[width=90 mm]{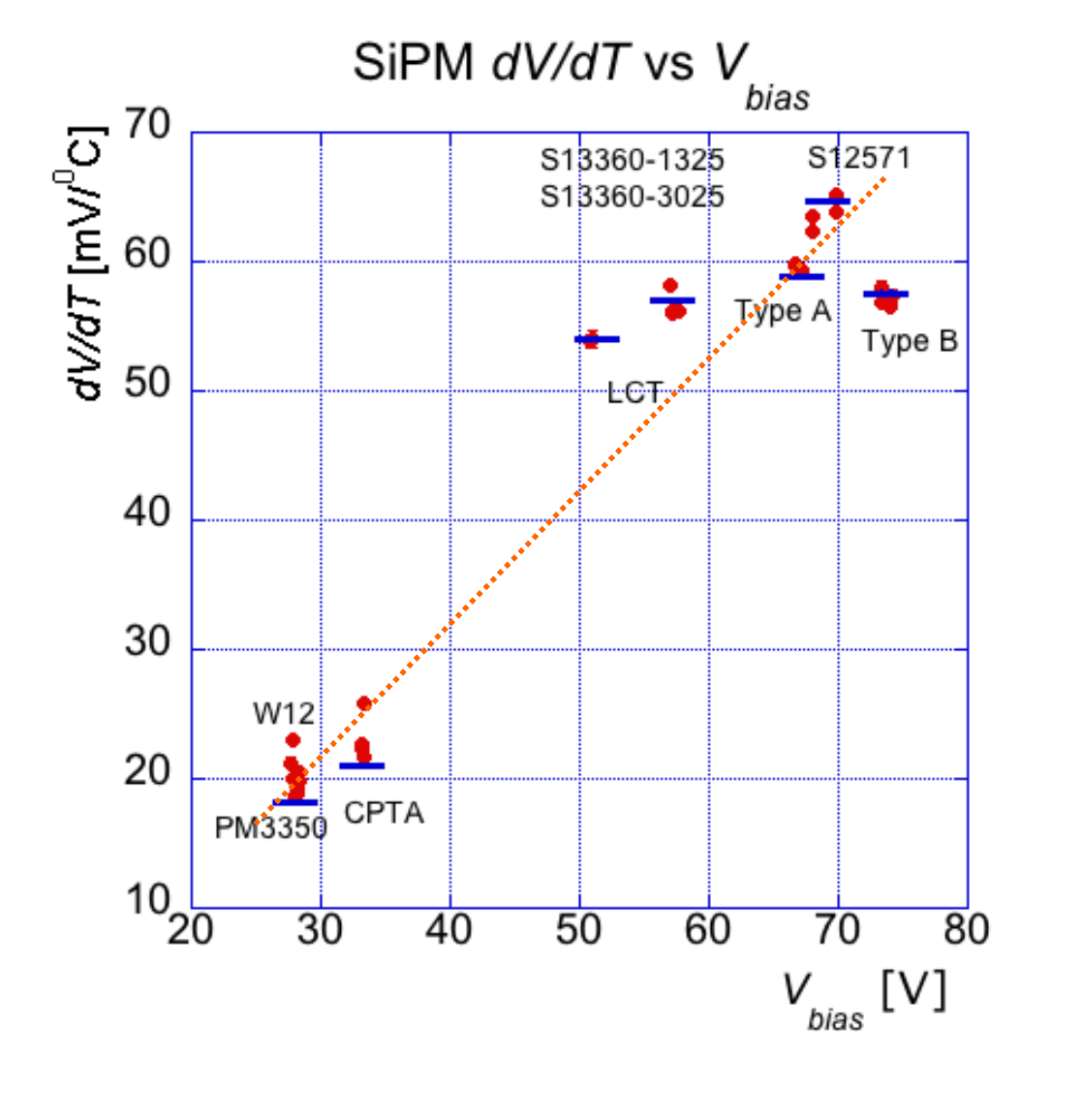}
\caption{Correlation of $dV_{\rm b}/dT$ versus bias voltage.}
\label{fig:dvdt-vbias}
\end{figure}

\subsection{ Gain Stabilization Results}

Table~\ref{tab:non-uniformity} summarizes the performance of the gain stabilization study for all 30 SiPMs. We succeeded in stabilizing 22 SiPMs including all 18 Hamamatsu MPPCs,  three CPTA SiPMs and one KETEK SiPM. 
Figure~\ref{fig:dvdt-vbias} shows the correlation of $dV_{\rm b}/dT$ versus the nominal  bias voltage. Note that Hamamatsu A-type  and S12571 sensors, KETEK PM3350 sensors and CPTA sensors lie on one line. The new MPPCs with trenches lie above the line as well as the experimental W12 SiPMs whereas the experimental B-type sensors lie below the line. This indicates that  for standard SiPMs the slope between $dV_{\rm b}/dT$ and $V_{b}$ is constant. Hamamatsu MPPCs with trenches operate at a lower bias voltage  than those without trenches but have a similar $dV_{\rm b}/dT$.
Typically the $dV_{\rm b}/dT$ relative  spreads for KETEK and CPTA SiPMs are larger than those for Hamamatsu MPPCs.

\begin{table}[htbp!]
\begin{center}
\caption{Measured gain deviations $\Delta G/G$ from uniformity in the $\rm 20^\circ - 30^\circ C$ temperature range.}
\begin{tabular}{|l|cccc|}  
\hline \hline
SiPM  & Ch1 $\Delta G /G$  & Ch2 $\Delta G/G$ &  Ch3 $\Delta G/G$ & Ch4 $\Delta G/G$   \\ \hline \hline
Hamamatsu  & A1-20:    & A2-20:   & A2-15:  & A1-15:  \\
A& $\pm 0.006\%$   & $\pm 0.029\%$  &  $\pm 0.009\%$  &  $\pm 0.007\%$  \\
& $(\pm 0.002\%)$  &  $(\pm 0.031\%)$ &   $(\pm 0.093\%)$&  $(\pm 0.042\%)$ \\\hline
Hamamatsu  & B1-20:  & B2-20: & B2-15:  & B1:15:  \\
 B&  $\pm 0.265\%$   &$\pm 0.079\%$ & $\pm 0.208\%$ & $\pm 0.325\%$   \\
 &  $(\pm 0.283\%)$ &$(\pm 0.127\%)$&  $(\pm 0.230\%)$ &  $(\pm 0.308\%)$ \\\hline
Hamamatsu  & S12571-271:  & S12571-273:  & S12571-137: & S12571-136: \\
 S12571-& $\pm 0.152\%$ &  $\pm 0.009\%$  & $\pm 0.009\%$  &  $\pm 0.271\%$ \\
 & $(\pm 0.195\%)$& $(\pm 0.012\%)$ & $(\pm 0.289\%)$  &  $(\pm 0.0210\%)$\\ \hline
Hamamatsu  &1325-10143: & 1325-10144:  & 3025-10104:  & 3025-10103:  \\ 
S13360-& $\pm 0.281\%$ &  $\pm 0.136\%$ &  $\pm 0.220\%$ & $\pm 0.185\%$ \\ \hline
Hamamatsu & LCT4\#6:       &   LCT4\#9:  & & \\   
&   $\pm  0.051  \%$       &   $ \pm 0.045 \%$ & & \\    \hline
CPTA & $\#857$:  & $\#922$: & $\#975$:  &$ \#1065$:  \\ 
& $\pm 0.539\%$ &  $\pm 0.197\%$ &  $\pm 0.267\%$ & $\pm 0.100\%$ \\ \hline
KETEK & W12-A:  & W12-B:  & PM3350-1:  &  PM3350-2: \\
 W12/PM3350& $\pm 0.381\%$ &  $\pm 1.361\%$ &  $\pm 1.943\%$ &   $\pm 1.685\%$ \\\hline
KETEK & PM3350-5:  & PM3350-6: &  PM3350-7:  &  PM3350-8: \\
 PM3350&$\pm 1.538\%$ &  $\pm 1.491\%$ &   $\pm 2.060\%$ &   $\pm 1.810\%$  \\
\hline \hline
\end{tabular}
\vskip -0.2cm
\label{tab:non-uniformity}
\end{center}
\end{table}

\section{Afterpulsing Effects}

\begin{figure}
\centering 
\includegraphics[width=150 mm]{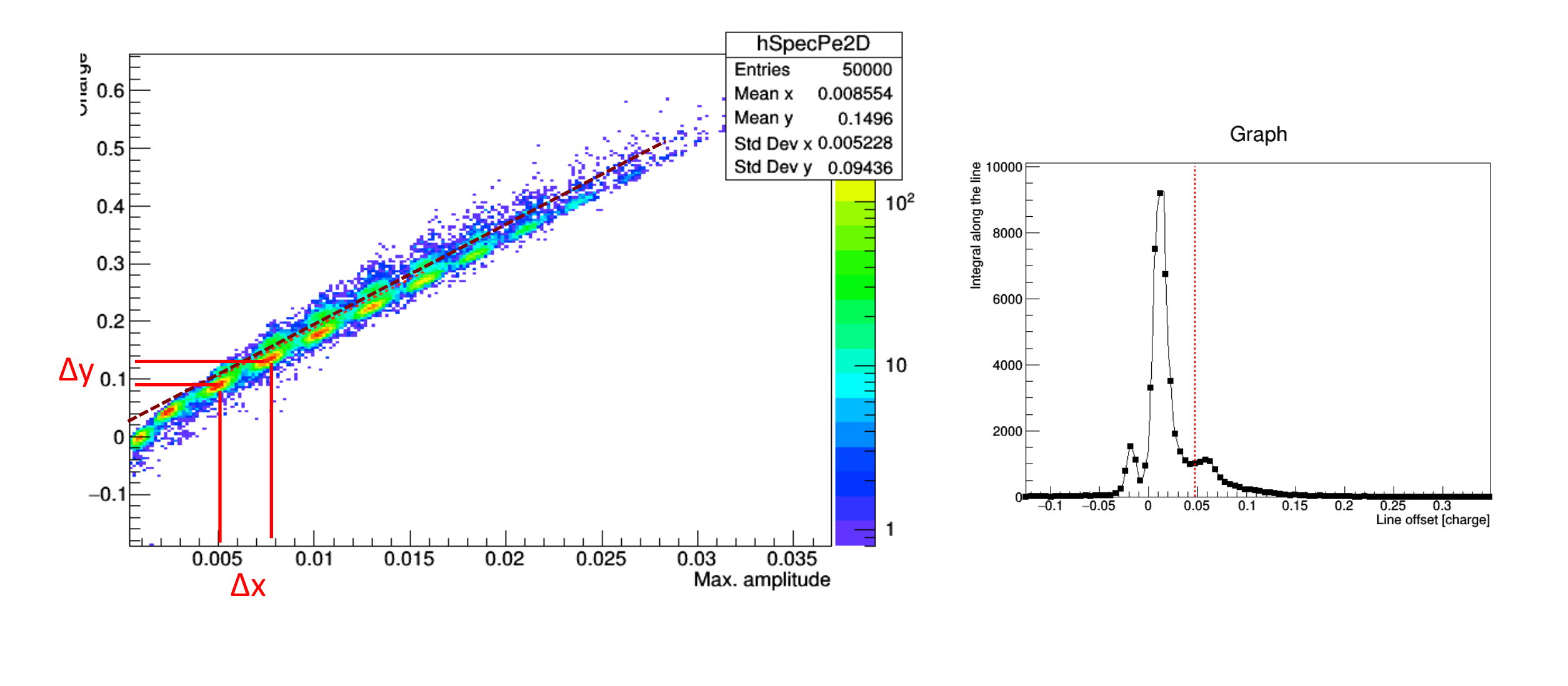}
\caption{Scatter plot of the integrated charge of the waveform versus the amplitude of the waveform minimum $A_{peak}$ (left) and the projection orthogonal to the dashed line (right). The peak represents all contributions of photoelectron peaks that are not affected by afterpulsing while the small peak to the right shows photoelectron peaks that are affected by afterpulsing. }
\label{fig:scatterplot}
\end{figure}

\begin{figure}
\centering 
\includegraphics[width=120 mm]{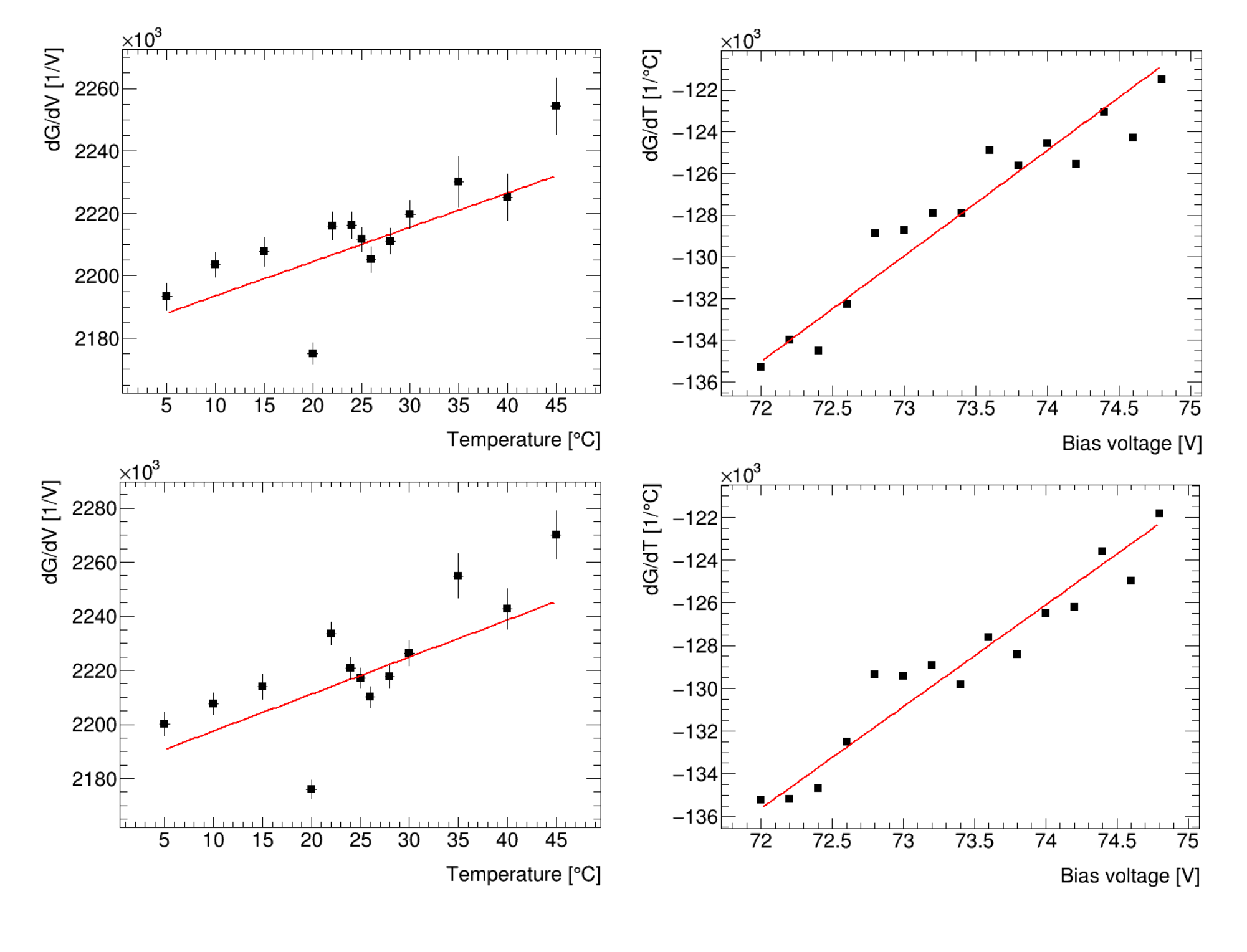}
\caption{The distributions of $dG/dV_{\rm b}$ versus $T$ for all data (top left), $dG/dT$ versus $V_{b}$  for all data (top right), $dG/dV_{\rm b}$ versus $T$ for reduced afterpulsing (bottom left) and $dG/dT$ versus $V_{b}$  for reduced afterpulsing (bottom right). }
\label{fig:afterpulsing}
\end{figure}

We developed two procedures to extract the photoelectron spectra from the measured waveforms, either by integrating the charge $Q$ or from  evaluating the amplitude of the waveform at the waveform minimum $A_{peak}$ position. The former is sensitive to contributions from afterpulsing while the latter is typically not. 
Thus, we can determine the amount of afterpulsing  from the scatter plot of $Q$ versu $A_{peak}$ plotted in Fig.~\ref{fig:scatterplot} (left). The red elliptical spots show the individual photoelectron peaks without afterpulse contributions lying on a diagonal. For waveforms with afterpulse contributions, $Q$ is shifted vertically since the waveform is broadened by the delayed second signal producing small satellite peaks separated from the peaks without afterpulsing by a valley. This becomes clearly visible if we plot the projection across the scatter plot shown in Fig.~\ref{fig:scatterplot} (right). Thus, we place a line at the valley position with a slope $a =\Delta y/ \Delta x$ that is determined from the separations ($\Delta x, \Delta y$) of the two-photoelectron peak and the three-photoelectron peak in the two observable $A_{peak}$ and $Q$, respectively.  If we select only waveforms that lie below the dashed line we obtain a sample with reduced afterpulsing. We have repeated the determination of $dG/dT$ and $dG/dV_{\rm b}$ for this sample with reduced afterpulsing. 
Figure~\ref{fig:afterpulsing} shows $dG/dV_{\rm b}$ versus $T$ and $dG/dT$ versus $V_{\rm b}$  for all waveforms and those with reduced afterpulsing. 
The $dG/dV_{\rm b}$ and $dG/dT$ distributions for waveforms with reduced afterpulsing look rather similar as those for all waveforms.
 Within errors we obtain the same $dG/dV_{\rm b}$ and $dG/dT$ for both samples indicating that afterpulsing has no effect on the gain stability studies.

\begin{figure}
\centering 
\includegraphics[width=120 mm]{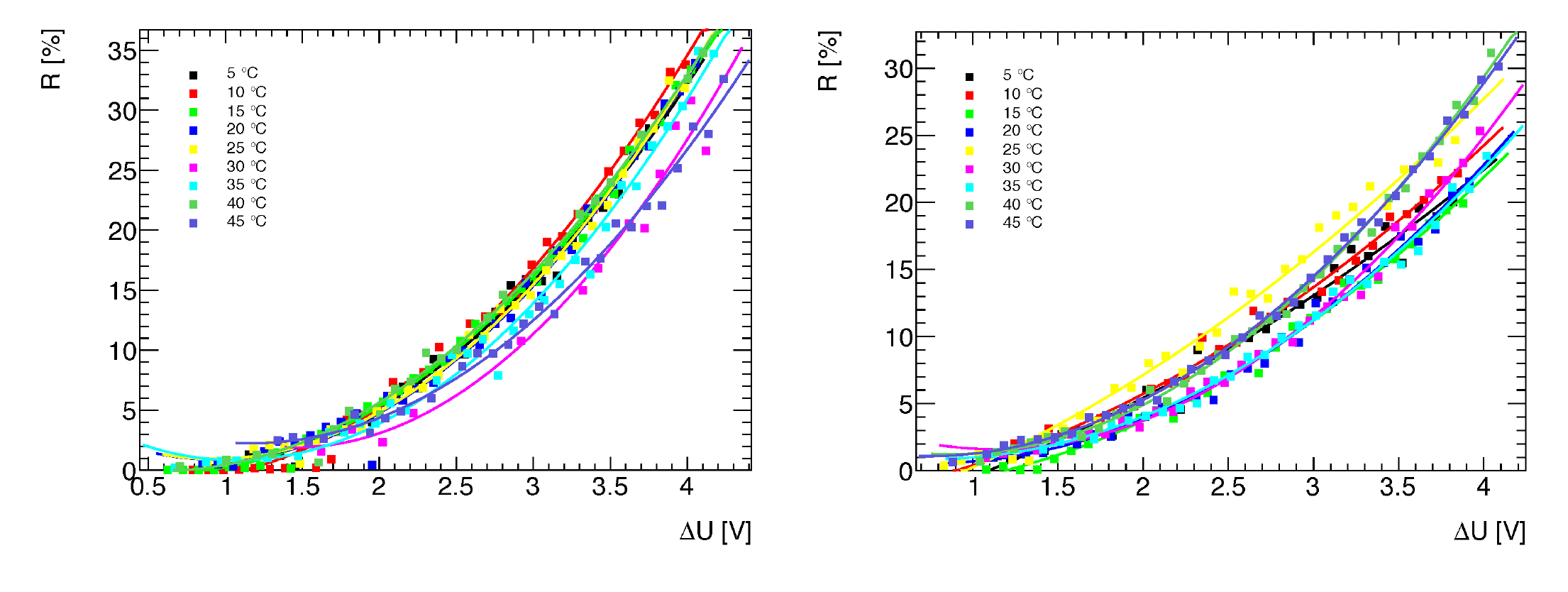}
\caption{The fraction aflerpulse waveforms as a function of overvoltage  for different temperatures for LCT4\#6 (left) and LCT4\#9 (right). }
\label{fig:LCT}
\end{figure}

We define the ratio of afterpulse waveforms $R$ as the entries that lie above the  red dashed line in Fig.~\ref{fig:scatterplot} to all entries and study $R$ as a function $V_{b}$ and $T$. Figure~\ref{fig:LCT} shows $R$ versus  $V_{b}$ for different temperatures for both LCT MPPCs. The fraction of afterpulse waveforms increase strongly with the overvoltage $\Delta U$. For $\Delta U =1~\rm V$, $R$ is less than $1\%$ while for $\Delta U =4~\rm V$, $R$ increase to $> 30\%$. 
We observe no explicit temperature dependence. The spread in the different curves indicates the systematic effect of the procedure. 

\section{Conclusions}     

We successfully completed gain stabilization tests for 30 SiPMs and demonstrated that batches of SiPMs can be stabilized with one $dV_{\rm b}/dT$ correction.
All 18 Hamamatsu MPPCs, six with trenches and 12 without trenches, satisfy the goal of $\Delta G/G < \pm 0.5\%$ in the $\rm 20^\circ C- 30^\circ C$
temperature range  many  MPPCs satisfy this in the extended temperature range of $\rm 1^\circ C- 48^\circ C$. Gain stabilization of KETEK SiPMs is more complicated since  the signals are rather long and are  affected by afterpulsing. The temperature range is limited to  $\rm 1^\circ C- 30^\circ C$. We succeeded in stabilizing only one out of the eight SiPMs tested. The $V(T)$ behavior is more complicated  requiring  individual $dV_{\rm b}/dT$ values to stabilize the gain of four SiPMs in the  $\rm 20^\circ C- 30^\circ C$ temperature range. Gain stabilization of CPTA SiPMs works fine. Three out of the four SiPM satisfy our criterion despite the fact that the LED light had to be absorbed by the scintillator and/or wavelength-shifting fiber before reaching the SiPM. Thus, this demonstrates that our procedure is applicable to a full tile/SiPM setup. 
We checked all Hamamatsu MPPCs without trenches with new fit methodology and obtain consistent results. For  MPPCs with trenches we need another function to represent tail on the right-hand side of each photoelectron peak. In the analog HCAL for ILC, the bias voltage  adjustment will be implemented on electronics board. Gain stabilization looks promising if the temperature is well measured and SiPM with similar properties are stabilzed with one $dV_{\rm b}/dT$ correction. Afterpulsing does not affect the gain stabilization results and  afterpulsing depends strongly  on overvoltage but not on temperature.

\section{Acknowledgments}  

This  work  was  conducted  in  the  framework  of  the  European  network  AIDA2020.   It  has  been
supported by the Norwegian Research Council and by the Ministry of Education, Youth and Sports
of the Czech Republic under the project LG14033.We would like to thank L. Linssen, Ch. Joram, W. Klempt, and D. Dannheim  for using the E-lab and for supplying electronic equipment. We further would like to thank the team of the climate chamber at CERN for their  assistance and support.

\eject

\section{ Appendix A}

\begin{figure}[htbp!]
\centering 
\includegraphics[width=120 mm]{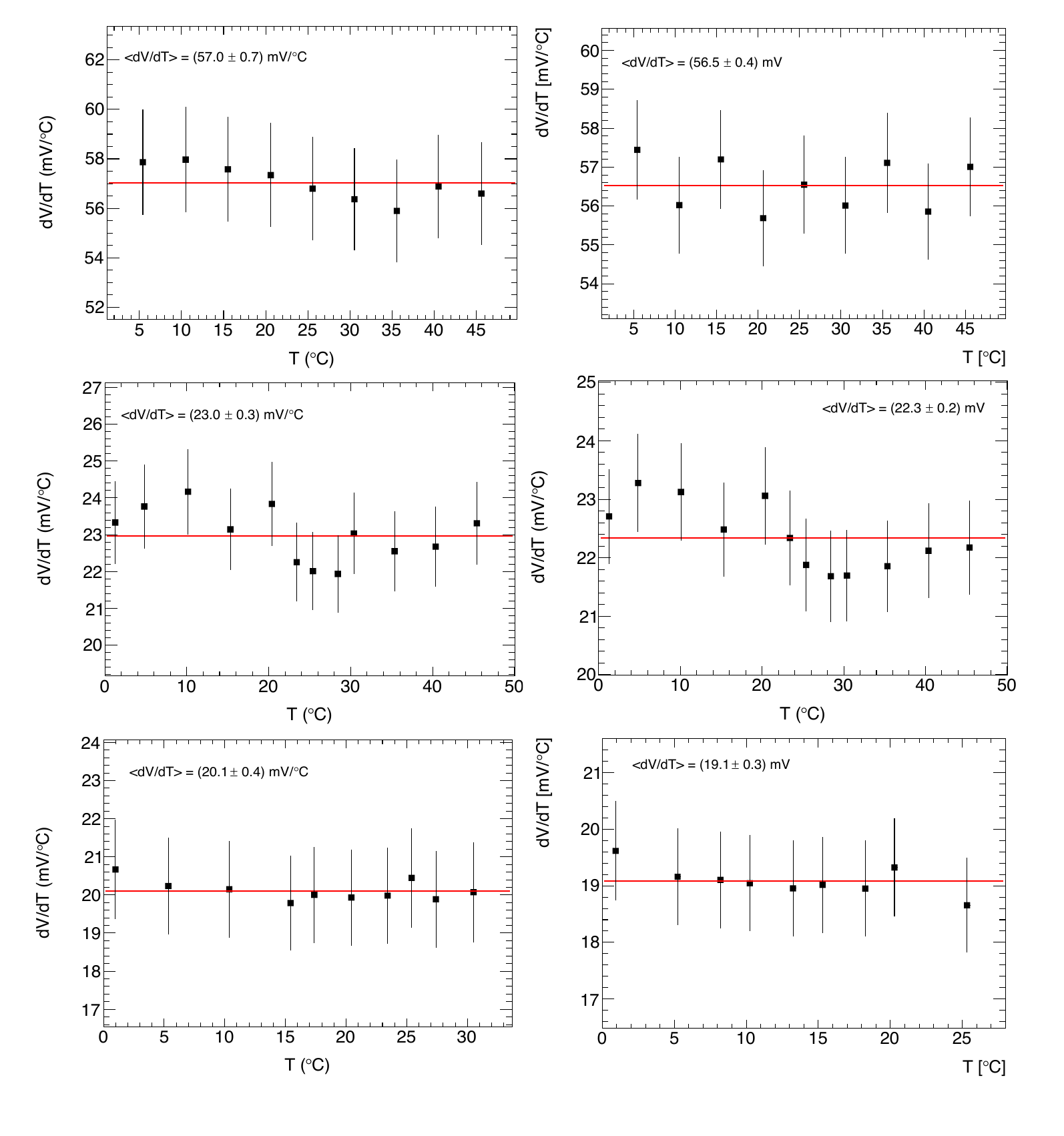}
\caption{\dvdt\ results for Hamamatsu B2-20 for new fit methodology (top left) and original fit methodology (top right), for CPTA-\#1065 for new fit methodology (middle left) and original fit methodology (middle right) and for KETEK PM3350-6 for new fit methodology (bottom left) and original fit methodology (bottom right).}
\label{fig:comparefits}
\end{figure}

\begin{table}
\begin{center}
\caption{Properties of tested SiPMs. For Hamamatsu and KETEK SiPMs, operating voltage and gain are specified for 25$^{\circ}$C, while for
CPTA SiPMs  the specifications are for 22$^{\circ}$C.}
\label{tab:sipm}
\begin{tabular}{|l|c|c|c|c|c|}
\hline \hline
\textbf{SiPM} & \textbf{Sensitive } &  \textbf{Pixel pitch} &  \textbf{$\#$pixels} & \textbf{Nominal} & \textbf{Typical} \\
\textbf{Type $\#$} & area \textbf{$[mm^2]$} &  \textbf{$[{\mu m} ]$} &  & $V_{\rm b}$ [V] &  \textbf{ $G~ [\times 10^5] $} \\
 [0.7ex]
 \hline
Hamamatsu & & & & & \\
A1  & $1 \times 1$ & $20$ & 2500 & $66.7$ & $2.3$ \\
A1  & $1 \times 1$ & $15$ & 4440 & $67.2$ & $2.0$ \\
A2  & $1 \times 1$ & $20$ & 2500 & $66.7$ & $2.3$ \\
A2  & $1 \times 1$ & $15$ & 4440 & $67.2$ & $2.0$ \\

B1  & $1 \times 1$ & $20$ & 2500 & $73.3$ & $2.3$ \\
B1  & $1 \times 1$ & $15$ & 4440 & $74.2$ & $2.0$ \\
B2 & $1 \times 1$ & $20$ & 2500 & $73.4$ & $2.3$ \\
B2  & $1 \times 1$ & $15$ & 4440 & $74.0$ & $2.0$ \\

S13360-1325-10143  & $1.3 \times 1.3$ & $25$ & 2668 & $57.2$ & $7.0$ \\
S13360-1325-10144  & $1.3 \times 1.3$ & $25$ & 2668 & $57.1$ & $7.0$ \\
S13360-3025-10103  & $3 \times 3$ & $25$ & 14400 & $57.7$ & $7.0$ \\
S13360-3025-10104  & $3 \times 3$ & $25$ & 14400 & $57.0$ & $7.0$ \\

LCT4\#6  & $1 \times 1$ & $50$ & 400 & $50.8$ & $16.0$ \\
LCT4\#9  & $1 \times 1$ & $50$ & 400 & $51.0$ & $16.0$ \\

S12571-010-271  & $1 \times 1$ & $10$ & 10000 & $69.8$ & $1.4$ \\
S12571-010-272  & $1 \times 1$ & $10$ & 10000 & $69.9$ & $1.4$  \\
S12571-015-136  & $1 \times 1$ & $15$ & 4489 & $68.1$ & $2.3$ \\
S12571-015137  & $1 \times 1$ & $15$ & 4489 & $68.0$ & $2.3$  \\
 [0.7ex]
 \hline
CPTA  & & & & & 
\\
\#857 & $1 \times 1$ & $40$ & 796 & 33.4 & 7.1 \\
\#922 & $1 \times 1$ & $40$ & 796 & 33.1 & 6.3 \\
\#975 & $1 \times 1$ & $40$ & 796 & 33.3 & 6.3  \\
\#1065 & $1 \times 1$ & $40$ & 796 & 33.1 & 7.0 \\ 
 [0.7ex]
\hline
KETEK  & & & & & 
\\
W12A & $3 \times 3$ & 20 & 12100 & 28 & 5.4 \\
W12B & $3 \times 3$ & 20 & 12100 & 28 & 5.4 \\
PM3350-1 & $3 \times 3$ & 50 & 3600 & 29.5 & 20 \\
PM3350-2 & $3 \times 3$ & 50 & 3600 & 29.5 & 20 \\
PM3350-5 & $3 \times 3$ & 50 & 3600 & 29.5 & 20 \\
PM3350-6 & $3 \times 3$ & 50 & 3600 & 29.5 & 20 \\
PM3350-7 & $3 \times 3$ & 50 & 3600 & 29.5 & 20 \\
PM3350-8 & $3 \times 3$ & 50 & 3600 & 29.5 & 20 \\
\hline \hline

\end{tabular}
\end{center}
\end{table}

\begin{table}[htbp!]
\begin{center}
\caption{Measured $dV_{\rm b}/dT$ values using the orginal fit model.}
\begin{tabular}{|l|cccc|}  
\hline \hline
SiPM type & Ch1 $dV_{\rm b}/dT$  & Ch2 $dV_{\rm b}/dT$ &  Ch3 $dV_{\rm b}/dT ]$ & Ch4 $dV_{\rm b}/dT $   \\ 
&  $[\rm mV/^\circ C]$  & $[\rm mV/^\circ C]$ &  $[\rm mV/^\circ C]$ &  $[\rm mV/^\circ C]$   \\ \hline \hline
Hamamatsu  & A1-20:    & A2-20:  & A2-15:  & A1-15:  \\
                       & $59.6\pm 0.4$  & $59.8\pm 0.3$  &  $59.3\pm 0.3$ & $59.2\pm 0.4$ \\ \hline
Hamamatsu  & B1-20:  & B2-20:  & B2-15: & B1-15:  \\
 &  $56.9\pm0.4$ & $58.0\pm 0.5$ & $56.5\pm 0.3$ &  $56.9\pm 0.4$ \\ \hline
Hamamatsu  & S12571-271:  & S12571-273: & S12571-137:  & S12571-136: \\
 & $63.9\pm 0.2$ &  $65.2\pm 0.2$ &  $62.3\pm 0.3$ &  $63.5\pm 0.3$ \\ \hline
Hamamatsu S13360- &1325-10143:  & 1325-10144:  & 3025-10104:  & 3025-10103:\\
 S13360- &$56.2\pm 0.3$ &  $58.1\pm 0.3$ &  $56.1 \pm 0.1$ &  $56.0\pm 0.2$\\ \hline
Hamamatsu & LCT$4\#6$:  & LCT$4\#9$:  &- & - \\ 
& $53.9\pm 0.5$ & $54.0\pm 0.7$ &- & - \\ \hline
CPTA & \#857:  & \#922:  & \#975:  & \#1065:  \\ 
 &  $21.6\pm 0.4$ & $22.5\pm 0.2$ &  $25.9\pm 0.3$ &  $22.3\pm 0.2$ \\ \hline
KETEK & W12-A:  & W12-B:  & PM3350-1:  &  PM3350-2:  \\
& $21.2\pm 0.4$ & $23.0\pm 0.2$ &  $20.0\pm 0.3$ &  $18.7\pm 0.4$ \\ \hline
KETEK  &PM3350-5:  & PM3350-6:   &  PM3350-7:  &  PM3350-8:  \\
 & $18.8\pm 0.2$ &   $19.1\pm 0.3$ &   $20.5\pm 0.2$ &   $19.8\pm 0.4$  \\
\hline \hline
\end{tabular}
\vskip -0.2cm
\label{tab:stabilization}
\end{center}
\end{table}

\begin{table}[htbp!]
\begin{center}
\caption{Comparison of $dV_{\rm b}/dT$ values obtained with the new fit model and with the original fit model for three SiPMs.}
\begin{tabular}{|l|cc|}  
\hline \hline
SiPM &  $dV_{\rm b}/dT ~ [\rm mV/^\circ C]$ (new fit) &   $dV_{\rm b}/dT ~ [\rm mV/^\circ C]$ (original fit) \\ \hline
Hamamatsu B2-20 & $57.0\pm 0.7$ & $56.5 \pm 0.4$ \\
CPTA 1065 & $23.0\pm 0.3$ & $22.3 \pm 0.2$ \\
KETEK PM$3350-6$ & $20.1 \pm 0.4 $ & $19.1\pm 0.3$  \\ \hline \hline
\end{tabular}
\vskip -0.2cm
\label{tab:comparefits}
\end{center}
\end{table}

\end{document}

%% file: econfmacros.tex
\def\beq{\begin{equation}}
\def\eeq#1{\label{#1}\end{equation}}
\def\eeqn{\end{equation}}

\def\beqa{\begin{eqnarray}}
\def\eeqa#1{\label{#1}\end{eqnarray}}
\def\eeqan{\end{eqnarray}}

\let\bar=\overbar

\def\Dslash{\not{\hbox{\kern-4pt $D$}}}
\def\dslash{\not{\hbox{\kern-2pt $\del$}}}

\def\msb{{\bar{\ssstyle M \kern -1pt S}}}

